\documentclass[lettersize,journal]{IEEEtran}
\usepackage{amsmath,amsfonts,amssymb,mathtools} 
\usepackage{physics}
\usepackage{algorithmic}

\usepackage{array}
\usepackage{xfrac}
\usepackage[caption=false,font=normalsize,labelfont=sf,textfont=sf]{subfig}
\usepackage{textcomp}
\usepackage{stfloats}
\usepackage{url}
\usepackage{verbatim}
\usepackage{graphicx}
\usepackage[table]{xcolor}
\usepackage{arydshln}
\usepackage{hhline,colortbl}
\usepackage{color,soul}
\usepackage[colorlinks=true, urlcolor=blue, citecolor=blue, linkcolor=blue]{hyperref}
\def\BibTeX{{\rm B\kern-.05em{\sc i\kern-.025em b}\kern-.08em
    T\kern-.1667em\lower.7ex\hbox{E}\kern-.125emX}}
\usepackage{balance}
\newcommand{\etal}{\textit{et al.}}

\usepackage[normalem]{ulem}
\def \nobreakseq {\nobreak \hskip 0pt \hbox}
\newcommand{\source}{{This is an archival version of our paper. Please cite the published version:  \href{https://doi.org/10.1109/TMECH.2023.3253250}{https://doi.org/10.1109/TMECH.2023.3253250}}}

\makeatletter
\def\ps@IEEEtitlepagestyle{}
\title{A Physics-Based Hybrid Dynamical Model of Hysteresis in Polycrystalline Shape Memory Alloy Wire Transducers}

\author{Michele A. Mandolino,
        Dominik Scholtes,\\
        Francesco Ferrante,~\IEEEmembership{Senior Member,~IEEE,}
        and~Gianluca Rizzello,~\IEEEmembership{Member,~IEEE}
\thanks{Manuscript received Month Day, 2022; revised Month Day, 2022; accepted Month Day, 2022. Recommended by Technical Editor --- and Senior Editor ---.}
\thanks{Michele A. Mandolino, Dominik Scholtes, and Gianluca Rizzello are with the Department of Systems Engineering, Saarland University, 66123 Saarbr\"ucken, Germany
        {\tt\footnotesize \{michele.mandolino, dominik.scholtes, gianluca.rizzello\}@imsl.uni-saarland.de}}%
\thanks{Francesco Ferrante is with the Department of Engineering, University of Perugia, 06123 Perugia, Italy
        {\tt\footnotesize francesco.ferrante@unipg.it}}%
\thanks{Digital Object Identifier (DOI): ---.}}

\usepackage{fancyhdr}
\begin{document}
\maketitle

\begin{abstract}
Shape Memory Alloys (SMAs) are a class of smart materials that exhibit a macroscopic contraction of up to 5\% when heated via an electric current. This effect can be exploited for the development of novel unconventional actuators.
Despite having many features such as compactness, lightweight, and high energy density, commercial SMA wires are characterized by a highly nonlinear behavior, which manifests itself as a load-, temperature\nobreakseq{-,} and rate-dependent hysteresis exhibiting a complex shape and minor loops. Accurate modeling and compensation of such hysteresis are fundamental for the development of high-performance SMA applications. 
In this work, we propose a new dynamical model to describe the complex hysteresis of polycrystalline SMA wires. 
The approach is based on a reformulation of the M\"{u}ller-Achenbach-Seelecke model for uniaxial SMA wires within a hybrid dynamical framework. In this way, we can significantly reduce the numerical complexity and computation time without losing accuracy and physical interpretability. After describing the model, an extensive experimental validation campaign is carried out on a 75 $\mu$m diameter SMA wire specimen. The new hybrid model will pave the development of hybrid controllers and observers for SMA actuators.
\end{abstract}


\section{Introduction}
\IEEEPARstart{S}{HAPE} memory alloy (SMA) transducers commonly consist of NiTi wires which undergo a contraction in length when heated, e.g., via an electric current \cite{Chaudhari:2021,Langbein:2010}. The wire recovers its original shape once the current is removed, provided that it is preloaded with a mechanical biasing mechanism. SMA features include high energy density, high flexibility, bio-compatibility, actuation strain up to 5 \%, lightweight, and ability to simultaneously work as actuators and sensors (self-sensing) \cite{Ruth:2017}. 
SMAs have been used as mechatronic actuators in many fields, e.g., biomedical systems \cite{Zhong:2021,Petrini:2011}, artificial muscles \cite{Jeong:2021,Huang:2020}, robotics \cite{Sohn:2018}, automotive \cite{Sellitto:2019}, and aerospace \cite{Costanza:2020}.

One of the major challenges encountered when developing SMA actuators lies in the prediction and compensation of their complex temperature-, rate-, and load-dependent hysteresis. 
Various modeling approaches have been proposed for SMA in the literature \cite{Lexcellent:2013,Ciss:2016}. On the one hand, physics-based models offer a thermodynamically-consistent framework to describe SMA hysteresis and related physical phenomena \cite{Auricchio:2002,Lagoudas:2008,Stebner:2013,Chemisky:2014,Tshikwand:2022}.
These constitutive models can be effectively used to predict the structural response of complex structures coupled with SMA elements.
However, since those are generally implemented in finite element software, they suffer from high numerical complexity. As a further complication, finite element models are usually expressed in a mathematical formalism that differs from the state-space dynamic representation adopted in control engineering. Therefore, many of these models turn out to be unsuitable for model-based control applications in real-time. 
On the other hand, phenomenological models such based on Preisach and Prandtl-Ishlinskii operators have also been used to describe SMA devices. The ability of those models to reproduce the hysteresis minor loops in an accurate and efficient way, as well as the fact that they are naturally formulated in a control-oriented formalism, has made them highly popular in SMA hysteresis compensation applications \cite{Dutta:2005,Toledo:2017,Basaeri:2019,Yoong:2021,Joey:2021}. However, since those models lack physical interpretation, they are not able to predict how the SMA hysteresis changes in response to a different external temperature or applied mechanical load.

\begin{figure}[!t]
    \centering
    \includegraphics[width=7.8cm]{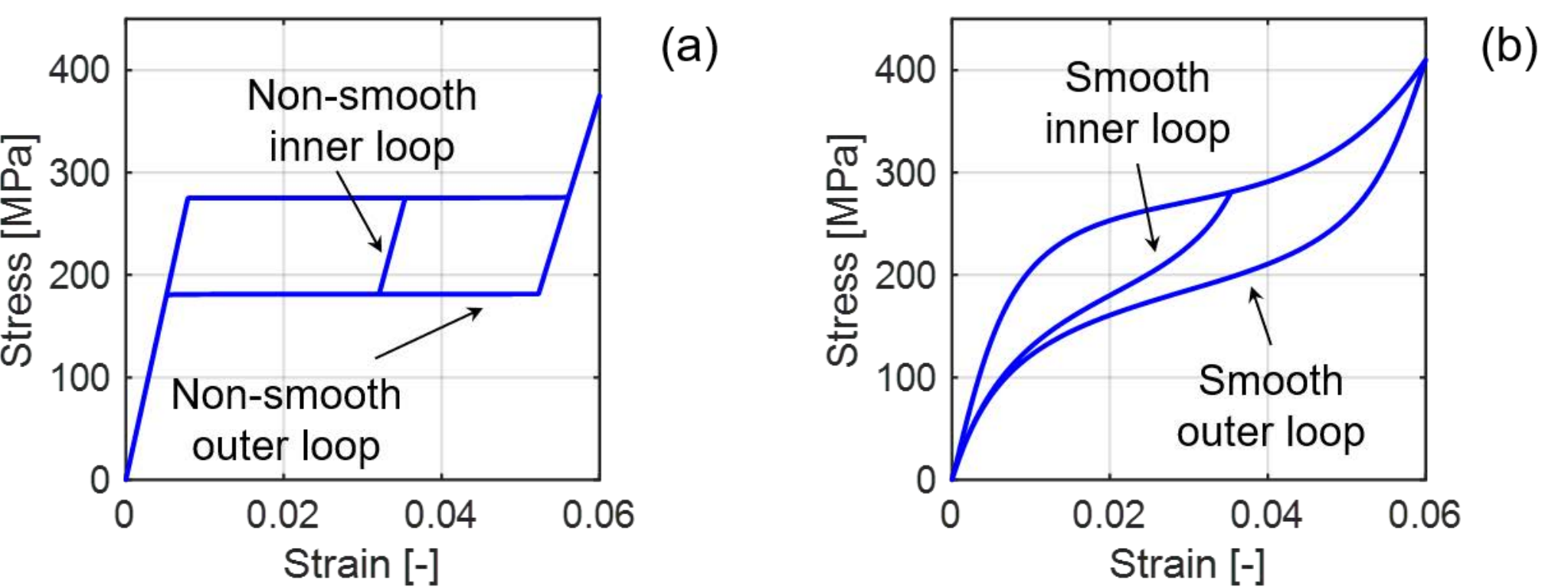}
    \caption{Example of single-crystal (a) and polycrystalline (b) SMA hysteresis.}
    \label{fig:SMA0}
\end{figure}

With the aim of exploiting the advantages of a physics-based SMA description within a control-oriented framework, in this work we propose a novel lumped-parameter dynamical model for one-dimensional polycrystalline SMA wire actuators.
Polycrystalline SMA wires, like the ones commonly available on the market, exhibit a complex hysteresis shape with minor loops upon partial loading/unloading.
The envisioned model must provide a numerically efficient and accurate tool to support SMA motion control applications in real-time. Ideally, the model must be formulated as a mechanical power port (i.e., an input-velocity, output-force state-space realization), so that it can be causally coupled with an external mechanical structure and, in turn, allow various types of SMA-driven systems to be described in a control-oriented fashion.
To achieve this goal, a valuable baseline is offered by the mesoscopic model for polycrystalline SMA wires presented by Rizzello \etal\: in \cite{Rizzello:2019}, which in turn is grounded on the physics-based and control-oriented description of single-crystal SMA provided by M\"{u}ller-Achenbach-Seelecke (MAS) \cite{Ballew:2019}.
In \cite{Rizzello:2019}, it is shown how such a model well reproduces smooth hysteresis loops, as well as minor loops, observed in a 508 $\mu$m superelastic SMA wire by NDC  as well as a 76 $\mu$m quasi-plastic wire by SAES Getters.
Despite those advantages, such a model is affected by high numerical stiffness due to strong nonlinearities, which results in large simulation times. 
A potential way to improve this limitation consists of eliminating the stiff dynamics from the model in \cite{Rizzello:2019} and replacing it with instantaneous hybrid transitions. This approach was already proven to be successful when modeling various electro-mechanical systems \cite{Ramirez-Laboreo:2019,Theunisse:2015}. The hybrid dynamical framework formalized in \cite{Goebel:2012} offers an ideal set of mathematical tools to obtain a convenient control-oriented model of the SMA. Other than providing advantages in terms of numerical robustness, simulation time, and sound mathematical modeling, this framework opens up the possibility to develop hybrid controllers for hysteresis compensation as well as hybrid observers for self-sensing applications \cite{Sanfelice:2020}, \cite{Bernard:2020}. At the same time, the hybrid framework allows to describe additional effects which are hard to take into account in a conventional modeling setting, i.e., the wire slack occurring upon loss of mechanical tension and intrinsic two-way effect. This approach is inspired by our previous work in \cite{Mandolino:2021}, where we developed a hybrid model for idealized single-crystal SMA hysteresis as in \figurename~\ref{fig:SMA0}(a). Here, the method from \cite{Mandolino:2021} is generalized for the first time to polycrystalline SMA exhibiting a smooth hysteresis and minor loops, as in \figurename~\ref{fig:SMA0}(b). The main challenges arise from the significantly higher complexity of the polycrystalline model compared to the single-crystal one, together with the need to find a suitable mathematical formulation of the slack dynamics. After presenting the hybrid model, an extensive experimental validation campaign is carried out on a 75 $\mu$m quasi-plastic  SMA wire by DYNALLOY. It is shown how the model predicts both the stress-strain and resistance-strain curves of the wire for different deformation rates and patterns, as well as applied electrical inputs. The new model also overperforms the one in \cite{Rizzello:2019} in terms of both simulation time and accuracy.

The remainder of this paper is organized as follows.
Section ~\ref{sec2} summarizes the physics-based MAS model for polycrystalline SMA wires. The novel hybrid dynamical model is then presented in Section~\ref{sec3}. In Section~\ref{sec4}, model characterization and validation are performed via a dedicated experimental setup. Concluding remarks are finally discussed in Section~\ref{sec5}.

\begin{figure}[!t]
    \centering
    \includegraphics[width=8.4cm]{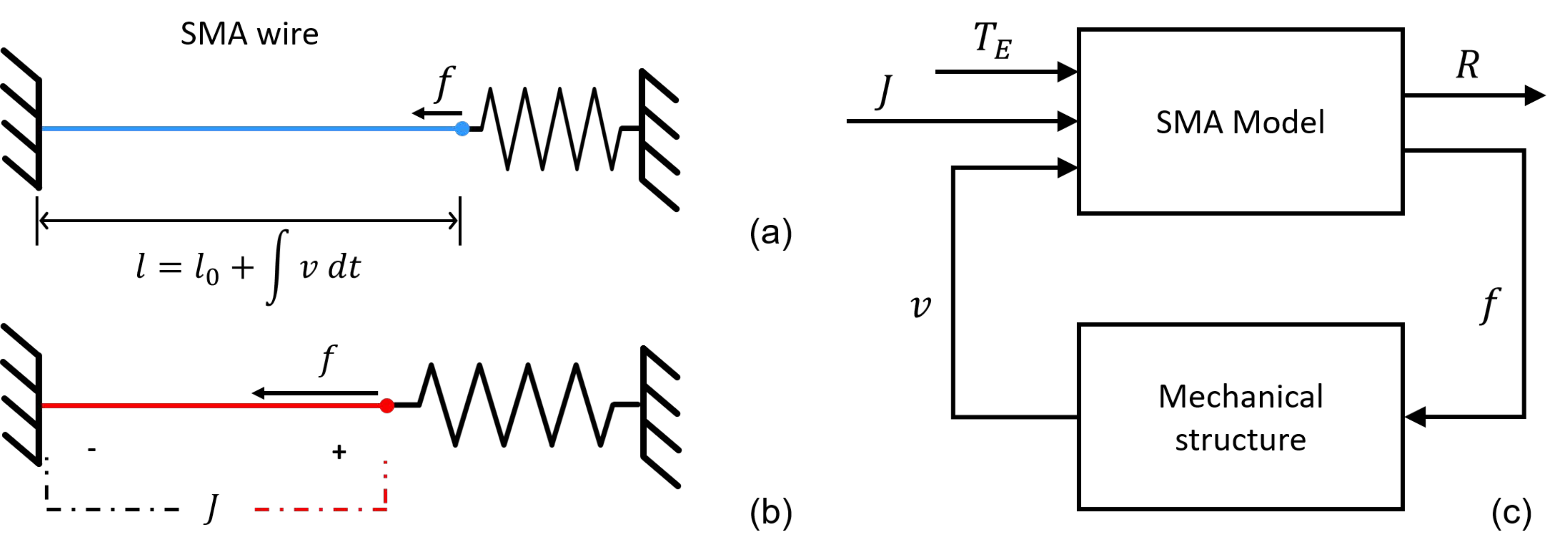}
    \caption{An unactuated SMA wire coupled to a mechanical bias structure (a). When we apply electric power to the wire, the Joule effect heats up the material, producing a contraction due to the interaction with the bias system (b). Causal coupling between the SMA model and an external structure (c).}
    \label{fig:SMA1}
\end{figure}

\section{SMA constitutive modeling} \label{sec2}
\noindent The polycrystalline SMA model previously developed in \cite{Rizzello:2019} is briefly summarized in this section.
We assume that the crystal lattice of a SMA material can be divided into different phases, or variants, each one associated with a specific geometry. 
For the particular case of a uniaxial SMA wire, these phases are called respectively austenite and martensite. The relative amount of each variant within the material depends on the thermo-mechanical loading conditions. To describe the lattice distribution, we introduce phase fraction variables \(x_A\) for austenite and \(x_M\) for martensite, respectively, such that
\begin{equation} \label{eq:MAS1}
    x_A + x_M = 1,\quad x_A\in[0,\,1], \quad x_M\in[0,\,1] \:.
\end{equation}
By exploiting (\ref{eq:MAS1}), we can define the stress-strain relationship: 
\begin{equation} \label{eq:MAS2}
    \sigma := \sigma(\varepsilon,x_M) =
    \frac{\varepsilon-\varepsilon_T x_M}{E_M^{-1} x_M + E_A^{-1}(1-x_M)} \:,
\end{equation}
where \(\sigma\) is the SMA axial stress, \(\varepsilon\) is the wire axial strain, \(E_A\) and \(E_M\) are austenite and martensite Young’s moduli respectively, while \(\varepsilon_T\) represent the transformation strain. 
The dependence of \(\sigma\) on phase fraction $x_M$ introduces a temperature-dependent stress-strain hysteresis.
Strain and stress can be related to wire force \(f\), length \(l\), and deformation rate \(v\) via
\begin{align}
    \label{eq:MAS3a}
    f =\pi r_0^2 \sigma \:,\quad
    l = l_0 (1 + \varepsilon) \:,\quad
    v = \dot{l} = l_0 \dot{\varepsilon} \:,
\end{align}
where \(r_0\) and \(l_0\) represent the cross-sectional radius and length of the undeformed and fully austenitic SMA wire, respectively. 
To predict the dynamic evolution of \(x_M\), we define
\begin{equation}\label{eq:MAS13}
    \dot{x}_M = -p_{MA}(\sigma,T)x_M + p_{AM}(\sigma,T)\left(1-x_M\right) \:,
\end{equation}
where \(p_{MA}\) and \(p_{AM}\) depend on macroscopic stress \(\sigma\) and temperature \(T\), and represent the probability of a mesoscopic martensite layer to transform into austenite and vice-versa. The generic expression of the transition probabilities is as follows
\begin{equation}\label{eq:MAS14}
    p_{\alpha \beta}(\sigma,T) = \tau_x^{-1} e^{-\frac{V_L}{k_B T}\Delta g_{\alpha \beta}(\sigma,T)} \:,
\end{equation}
where \(\tau_x\) is a time constant related to the thermal activation, \(V_L\) is the volume of a mesoscopic layer, \(k_B\) is the Boltzmann constant, and \(\Delta g_{\alpha \beta}(\sigma,T)\) represents the energy barrier existing between phases \(\alpha\) and \(\beta\) in the Gibbs free-energy density landscape. 
The evolution of the SMA temperature can be determined via the internal energy balance equation, namely
\begin{equation} \label{eq:MAS4}
    \Omega \rho_V c_V \dot{T} = - \lambda A_S (T-T_E) + J + \Omega \rho_V h_M \dot{x}_M \:.
\end{equation}
In \eqref{eq:MAS4}, \(\Omega=\pi r^2_0 l_0\) represents the wire volume, \(\rho_V\) is the SMA volumetric density, \(c_V\) is the specific heat, \(\lambda\) is the convective cooling coefficient between SMA and environment, \(A_S = 2 \pi r_0 l_0\) is the heat exchange lateral surface of the wire, \(T_E\) is the environmental temperature, \(J\) is the Joule heating generated by an applied electric current used to actuate the wire, and  \(h_M\) is the specific latent heat \cite{Rizzello:2019,Ballew:2019}.

\begin{figure}[!t]
    \centering
    \includegraphics[width=8.4cm]{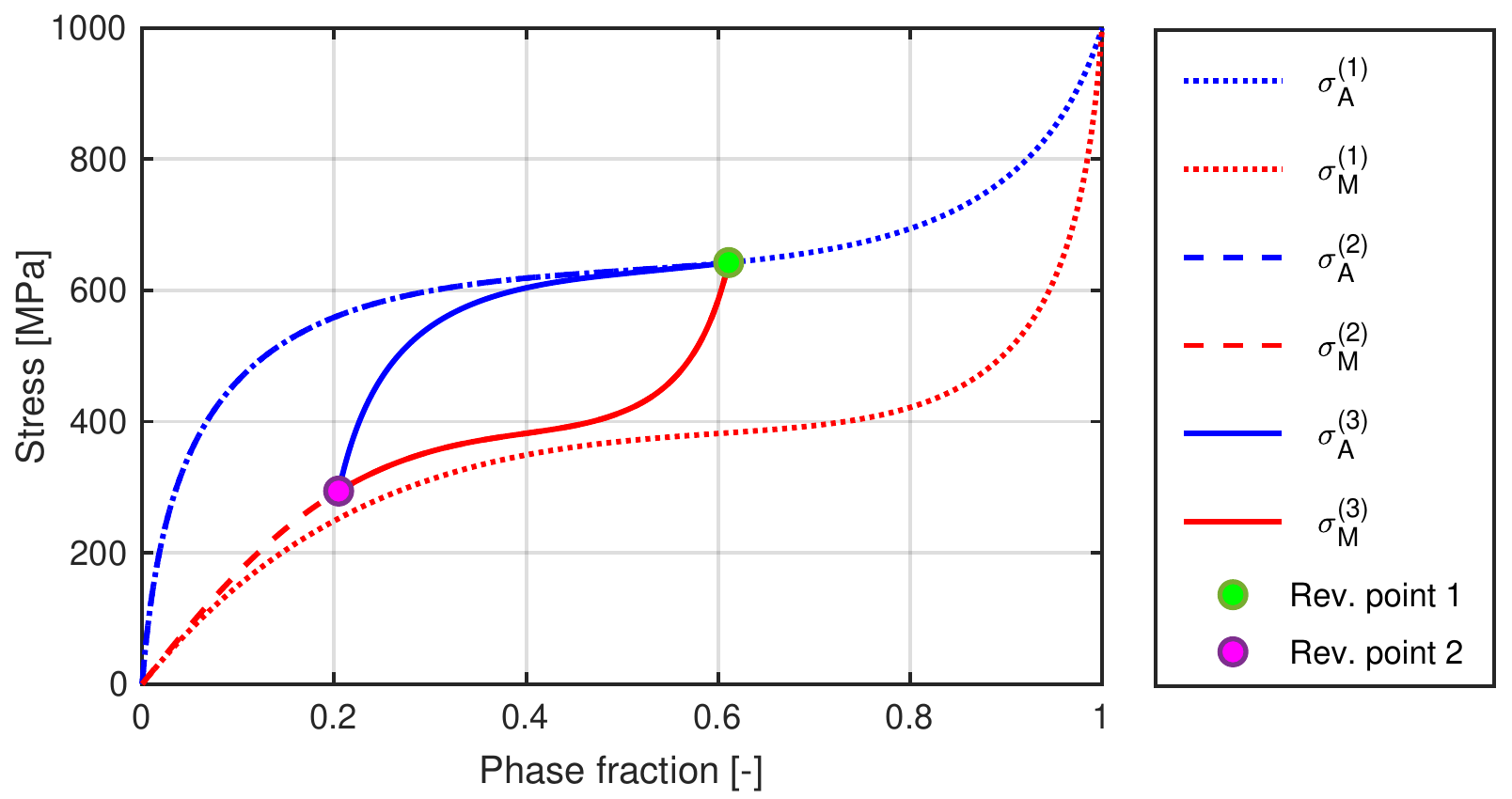}
    \caption{Example of SMA inner hysteresis loops in the stress-phase plane.}
    \label{fig:SMA0a}
\end{figure}

The definition of \(\Delta g_{\alpha\beta}(\sigma,T)\) in \eqref{eq:MAS14} depends on the SMA transformation stresses (see \cite{Rizzello:2019}  for details). 
Those quantities represent the upper and lower branches of the hysteresis minor loop in which we are currently moving. For the \(k\)-th nested minor hysteresis loop, those upper and lower hysteresis branches are denoted as \(\sigma_A^{(k)}\) and \(\sigma_M^{(k)}\), respectively, such that
\begin{align} \label{eq:MAS7_1}
    \sigma_A^{(k)}(x_M,T) : \mathbb{X}_M^{(k)} \times \mathbb{R}_{\geq 0} \rightarrow \mathbb{R} \:, \\
    \label{eq:MAS7_2}
    \sigma_M^{(k)}(x_M,T) : \mathbb{X}_M^{(k)} \times \mathbb{R}_{\geq 0} \rightarrow \mathbb{R} \:,
\end{align}
where \(\mathbb{X}_M^{(k)} = [\underline{x}^{(k)}_M, \bar{x}^{(k)}_M]\) represents the admissible range of \(x_M\) for the current inner loop \(k\).
For \(k=1\), \(\underline{x}^{(1)}_M = 0\), \(\bar{x}^{(1)}_M = 1\), and \(\sigma_A^{(1)}\) and \(\sigma_M^{(1)}\) denote the outermost hysteresis loop (cf. \figurename~\ref{fig:SMA0a}).
While transformation stresses were originally derived from stochastic distributions due to internal stresses and energy barriers \cite{Heintze:2008}, in practice they can be extracted from tensile experiments. 
Suitable outer loop interpolators are given by \cite{Rizzello:2019}
\begin{align} 
    \label{eq:MAS7a}
    \sigma_A^{(1)}(x_M,T) &= 
    \sigma_{A0}(x_M) + \sigma_{S}(x_M)(T-T_0) \:, \\
    \label{eq:MAS7b}
    \sigma_M^{(1)}(x_M,T) &= 
    \sigma_{M0}(x_M) + \sigma_{S}(x_M)(T-T_0) \:,
\end{align}
with
\begin{equation}
    \label{eq:MAS8}
    \begin{split}
  &\sigma_{A0}(x_M) = E_{AL} \log(1+\lambda_{AL}x_M) + \\
   &+ E_{AR} \log(1+\lambda_{AR}(1-x_M)) + E_{AC}x_M + \sigma_{AB} \:,
    \end{split}
\end{equation}
\begin{equation}
    \label{eq:MAS9}
    \begin{split}
    &\sigma_{M0}(x_M) = E_{ML} \log(1+\lambda_{ML}x_M) + \\
    &+ E_{MR} \log(1+\lambda_{MR}(1-x_M)) + E_{MC}x_M + \sigma_{MB} \:,
    \end{split}
\end{equation}
\begin{equation}
    \label{eq:MAS10}
    \begin{split}
    \sigma_{S}(x_M) = &\frac{E_{SL}}{1+e^{-\lambda_{SL}(x_M-x_{0SL})}} + \\
    &+  \frac{E_{SR}}{1+e^{\lambda_{SR}(x_M-x_{0SR})}} +E_{SC}x_M + \sigma_{SB} \:.
    \end{split}
\end{equation}
All parameters appearing in \eqref{eq:MAS8}-\eqref{eq:MAS10} represent constitutive material constants.
Once the outer hysteresis loop \(\sigma_A^{(1)}\) and \(\sigma_M^{(1)}\) is known, the model can predict an arbitrary inner hysteresis loop \(\sigma_A^{(k)}\) and \(\sigma_M^{(k)}\) for any \(k \in \mathbb{N}_{\geq 1}\), in response to a generic loading history. This prediction can be implemented by keeping in memory the past sequences of loading/unloading branches, namely \(\sigma_A^{(k)}\) and \(\sigma_M^{(k)}\) for \(k=1,\ldots,n_l\) where \(n_l\) denotes the current inner loop, as well as the reversal points corresponding to switches between loading/unloading, in terms of  \(\underline{x}^{(k)}_M\) and \(\bar{x}^{(k)}_M\). 
Whenever a reversal point occurs, the current inner loop \(n_l\) changes to \(n_l+1\) (see 
\figurename~\ref{fig:SMA0a}).
The new loading/unloading branches \(\sigma_A^{(n_l+1)}\) and \(\sigma_M^{(n_l+1)}\) are computed by scaling curves \(\sigma_A^{(n_l)}\) and \(\sigma_M^{(n_l)}\) accordingly. 
This process can be repeated recursively to compute any inner loop, regardless of its hierarchical level. Further policies are also implemented to account for inner loop closure, causing a change from \(n_l\) to \(n_l-2\). A detailed description of the theory
behind scaling policy and bookkeeping algorithm is beyond the scope of this paper, please refer to \cite{Rizzello:2019} for more details.

Finally, we consider a model for the SMA electrical resistance, which is important for self-sensing applications \cite{Ruth:2017}:
\begin{equation}\label{eq:MAS11}
    R = \frac{l_0 (1+\varepsilon)}{\pi r^2_0 (1-\nu \varepsilon)} \left[ \rho_{eM}(T)x_M + \rho_{eA}(T)\left(1-x_M\right) \right] \:,
\end{equation}
where \(\nu\) is the Poisson's ratio, \(\rho_{eM}\) and \(\rho_{eA}\) are the electrical resistivity of the two phase fraction of the material, given by
\begin{align}
    \label{eq:MAS12a}
    \rho_{eM} &= \rho_{eM}(T_0) [ 1 + \alpha_M(T-T_0) ] \:,\\
    \label{eq:MAS12b}
    \rho_{eA} &= \rho_{eA}(T_0) [ 1 + \alpha_A(T-T_0) ] \:.
\end{align}
Constants \(\rho_{eM}(T_0)\), \(\rho_{eA}(T_0)\), \(\alpha_M\), and \(\alpha_A\) in (\ref{eq:MAS12a})-(\ref{eq:MAS12b}) represent constitutive material parameters.

The complete model of the polycrystalline SMA wire can be obtained by collecting \eqref{eq:MAS2}-\eqref{eq:MAS13}, \eqref{eq:MAS4}, and \eqref{eq:MAS11}:
\begingroup\makeatletter\def\f@size{9.5}\check@mathfonts\def\maketag@@@#1{\hbox{\m@th\normalsize\normalfont#1}}%
\begin{equation} \label{eq:MAS15}
    \begin{cases}
        \dot{\varepsilon} = l_0^{-1}v \\[1em]
        \dot{x}_M \coloneqq \phi_{x_M} = -p_{MA}x_M + p_{AM}(1-x_M) \\[1em]
        \dot{T} = \dfrac{J - \lambda A_s(T-T_E)}{\Omega\rho_V c_V} + \dfrac{h_{M}\phi_{x_M}}{c_V}\\[1em]
        f = \pi r_0^2 \dfrac{\varepsilon - \varepsilon_T x_M}{E_M^{-1}x_M + E_A^{-1}(1 - x_M)}\\[1em]
        R = \dfrac{l_0 (1+\varepsilon)}{\pi r^2_0 (1-\nu \varepsilon)} \left[ \rho_{eM}(T)x_M + \rho_{eA}(T)\left(1-x_M\right) \right]
    \end{cases}.
\end{equation}\endgroup
The state variable of \eqref{eq:MAS15} are \(\varepsilon\), \(x_M\), and \(T\), the inputs are \(v\), \(J\), and \(T_E\), and the outputs are \(f\) and \(R\). A block diagram depiction of the model is shown in \figurename~\ref{fig:SMA1}(a). This representation allows us to express the SMA model in impedance form (velocity-input, force-output), so it can be easily coupled with a generic mechanical structure naturally expressed in admittance form in a causal way (e.g., a mass-spring-damper or Euler-Lagrange model, naturally expressed as force-input, velocity-output), see \figurename~\ref{fig:SMA1}(b). Therefore, model \eqref{eq:MAS15} allows to describe mechanical structures actuated by 1D SMA wires in a lumped-parameter and control-oriented fashion.

\section{SMA Hybrid Dynamical Model} \label{sec3}
\noindent The hybrid reformulation of \eqref{eq:MAS15} is discussed in this section.
\subsection{Preliminaries on Hybrid Systems} \label{sec:Hybrid}
\noindent We consider hybrid systems with state $x\in\mathbb{R}^n$ and input $u\in\mathbb{R}^m$ of the form
\begin{equation} \label{eq:HY1}
    \mathcal{H}\colon\left\lbrace
    \begin{array}{ccll}
        \dot{x} &  =  & F(x, u) & \quad (x, u)\in C\\
           x^+  & \in & G(x, u) & \quad (x, u)\in D
    \end{array},
    \right.
\end{equation}
where $F\colon\mathbb{R}^{n+m}
\rightarrow\mathbb{R}^n$ is the flow map, $C\subset\mathbb{R}^n$ is the flow set, $D\subset\mathbb{R}^n$ is the jump set, and the set-valued map $G\colon\mathbb{R}^n\rightrightarrows\mathbb{R}^n$ is the jump map. 
The symbol $\dot{x}$ denotes the time-derivative of state $x$ during flows, while $x^+$ represents the value of state $x$ after an instantaneous change.
To denote the above hybrid system, we use the shorthand notation $\mathcal{H}=(C, F, D, G)$. For more details on hybrid system, please refer to \cite{Goebel:2012}.

\subsection{Characterization of the Operative Modes} \label{sec:PC}

\begin{figure}[!t]
    \centering
    \includegraphics[width=8.4cm]{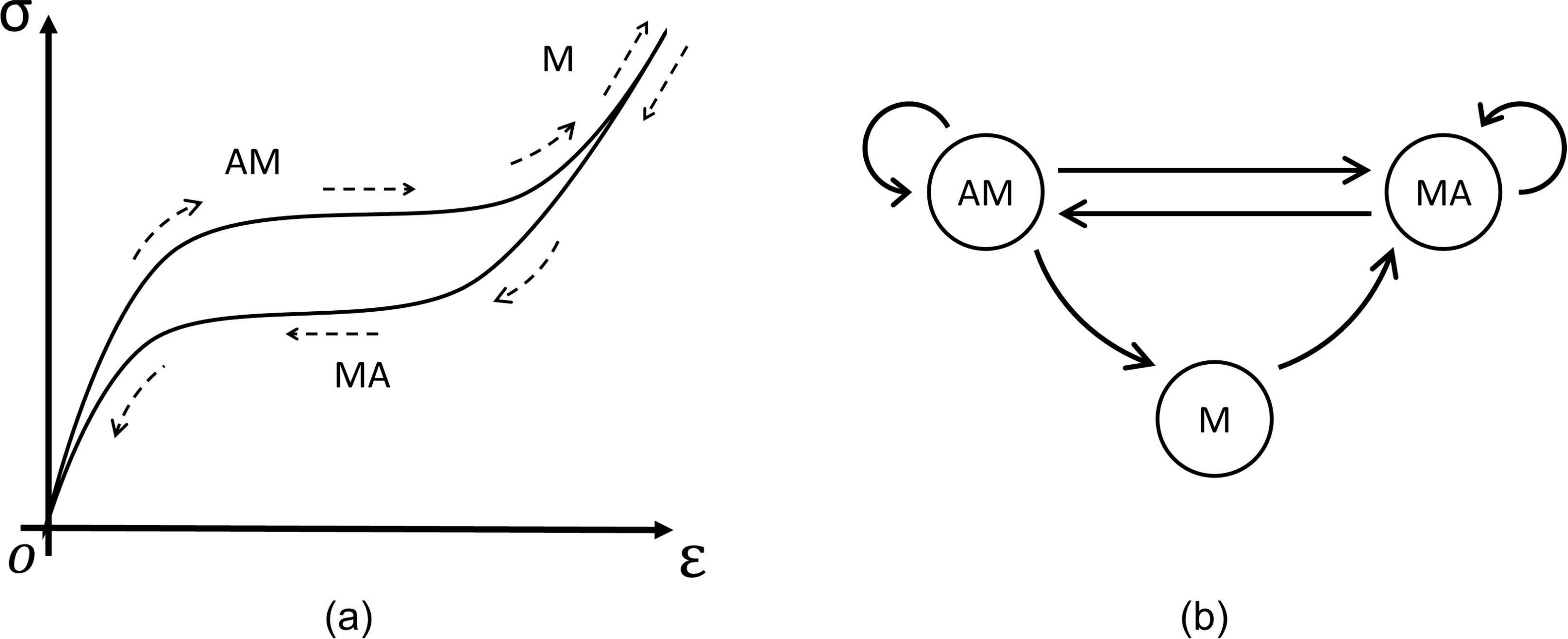}
    \caption{Qualitative sketch of a pseudoelastic stress-strain characteristic of a polycrystalline SMA wire (a), and corresponding finite state machine (b).}
    \label{fig:hybrid1}
\end{figure}

\noindent As pointed out in \cite[Assumption 2]{Rizzello:2018}, if we choose the values of \(\tau_x\) and \(V_L\) in a physically meaningful way, the transition probabilities \(p_{MA}\) and \(p_{MA}\) defined in (\ref{eq:MAS14}) behave approximately as high-gain threshold functions. 
As a result, \eqref{eq:MAS13} becomes responsible for the high numerical stiffness of model \eqref{eq:MAS15}.
Following the analysis in \cite{Rizzello:2018}, it can be shown that during phase transformation (i.e., \(\dot{x}_M \neq 0\)) the following approximation tightly holds for the polycrystalline model \eqref{eq:MAS15}
\begin{equation} \label{eq:HY2}
        \sigma(\varepsilon,x_M) = \sigma_A^{(n_l)}(x_M,T) \quad \text{if } \dot{x}_M > 0 \:,
\end{equation}
\begin{equation} \label{eq:HY2a}
        \sigma(\varepsilon,x_M) = \sigma_M^{(n_l)}(x_M,T) \quad \text{if } \dot{x}_M < 0 \:,
\end{equation}
where \(\sigma(\varepsilon,x_M)\) is defined in (\ref{eq:MAS2}), while \(\sigma_A^{(n_l)}(x_M,T)\) and \(\sigma_M^{(n_l)}(x_M,T)\) describe the current minor loop of the hysteresis. 

The hybrid reformulation is grounded on the MAS model structural property defined by \eqref{eq:HY2}-\eqref{eq:HY2a}.
Indeed, by using \eqref{eq:HY2}-\eqref{eq:HY2a}, we can compute \(x_M\) without the need to integrate the stiff equation \eqref{eq:MAS13}, therefore improving the numerical properties of model \eqref{eq:MAS15}, as shown in the sequel.
As a first step, a set of operative modes needs to be identified.
We consider in Fig.~\ref{fig:hybrid1}(a) the qualitative stress-strain hysteresis of a pseudoelastic polycrystalline SMA wire with \(T\) higher than the austenite transformation temperature \cite{Lagoudas:2008}.
When moving along the branches of this hysteresis, either \eqref{eq:HY2} or \eqref{eq:HY2a} always holds in a mutually exclusive manner. Based on this observation, we can determine three distinct operating modes:
\begin{enumerate}
    \item AM: Austenite to Martensite (or loading) branch;
    \item MA: Martensite to Austenite (or unloading) branch;
    \item M: full Martensite branch.
\end{enumerate}
The finite state machine which defines the transition logic between those modes is sketched in Fig.~\ref{fig:hybrid1}(b). A hypothetical operating sequence of the model is as follows. We assume that the pseudoelastic SMA wire starts in a full austenitic condition (mode AM). When subject to increasing mechanical load, the amount of austenitic crystal lattice is decreased while the martensitic one increases, i.e., \(\dot{x}_M > 0\) and thus \eqref{eq:HY2} holds while moving along the outer hysteresis loop. If the loading state exceeds a certain threshold (dictated by material-specific and external inputs), the SMA wire transforms completely into martensite (mode M). When the material is still working below the above threshold and we start to unload it, it gradually transforms from martensite to austenite (mode MA), i.e., \(\dot{x}_M < 0\) and thus \eqref{eq:HY2a} holds while moving along this inner hysteresis loop. Inner loops appearing upon several partial loading-unloading cycles are handled by the same AM and MA modes. For each of these transitions, a different pair of \(\sigma_A^{(n_l)}\) and \(\sigma_M^{(n_l)}\) is generated to describe the current minor loop, parameterized through variable $n_l$ (with $n_l = 1$ for the outer loop and $n_l > 1$ for the minor loops, respectively). 

\begin{figure}[!t]
    \centering
    \includegraphics[width=8.4cm]{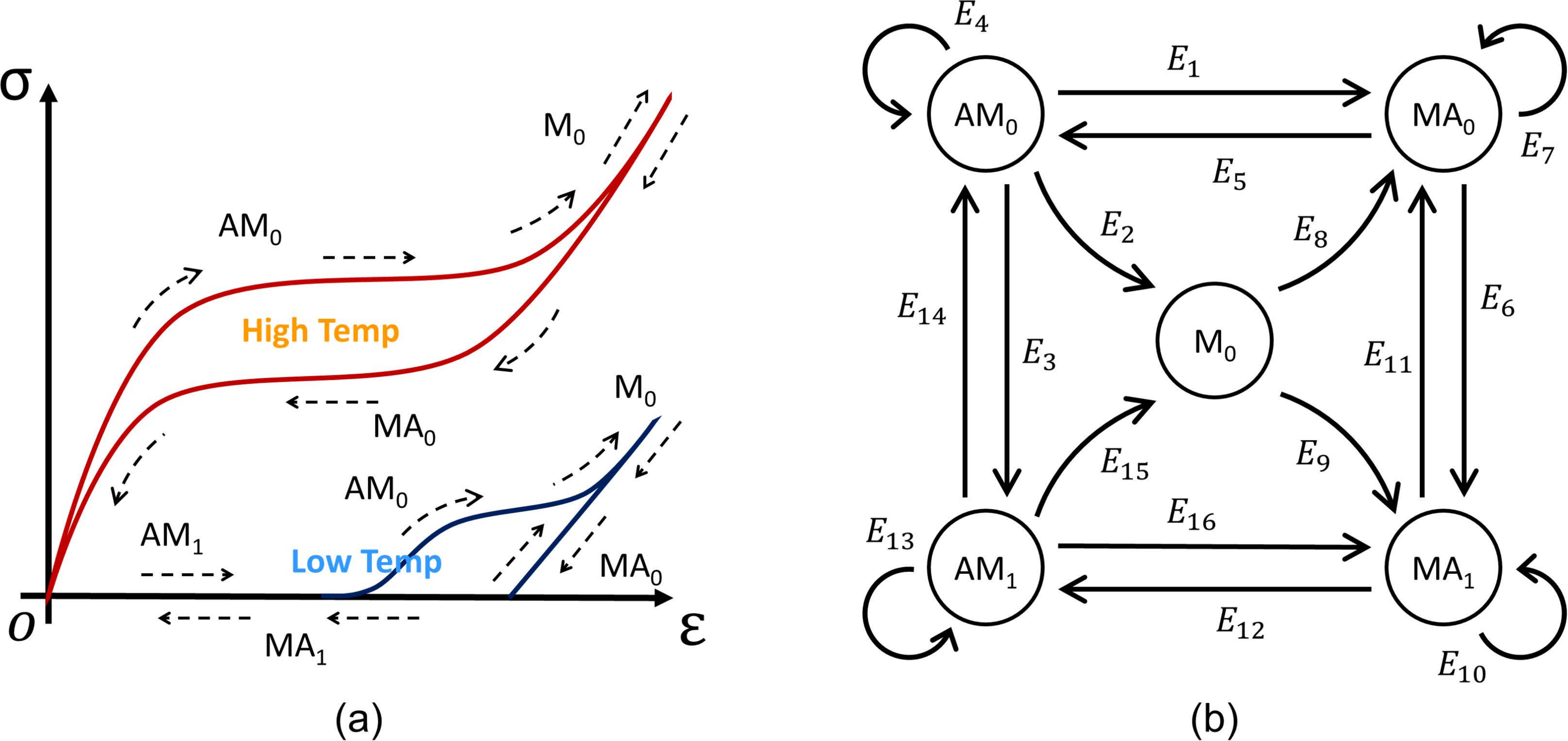}
    \caption{Different behaviors of a polycrystalline SMA wire at low and high temperatures (a), and the corresponding finite state machine which also includes the slack dynamics (b).}
    \label{fig:hybrid2}
\end{figure}

At room temperature, a SMA wire does not usually exhibit a stable austenitic phase. Instead, detwinning of the martensite causes a quasi-plastic material behavior, which results in a qualitative hysteresis shape as the blue one in Fig.~\ref{fig:hybrid2}(a). Compared to the pseudoelastic case, quasi-plastic SMA shows a region of zero stress. This corresponds to the wire being slack, and thus not subject to any mechanical tension, while still exhibiting a residual strain.
The slack condition is frequently observed when characterizing quasi-plastic SMA wires as well as in agonist-antagonist SMA actuator applications, in which the unactuated wire loses tension during normal operating conditions, only for regaining it after being re-activated \cite{Spaggiari:2013}.
This effect is not covered by model \eqref{eq:MAS15}, which becomes inaccurate when \(\sigma < 0\).
To model the slack, we include two additional modes by duplicating AM and MA. A subscript is attached to each mode to indicate whether the slack is present (1) or not (0). Note that the operative mode M is not duplicated, since it exists only for \(\sigma \geq 0\).
During slack the wire is no longer under tension, thus its axial stress becomes zero. By setting \(\sigma = 0\) in (\ref{eq:MAS2}), we obtain that \(\varepsilon = x_M \varepsilon_T\) always holds true in slack.
To quantify the SMA mechanical state in both tensioned and slacked conditions, we define an \textit{effective} (i.e., residual) strain \(\varepsilon_{eff}\) as
\begin{equation} \label{eq:HY8a}
    \varepsilon_{eff} \coloneqq \left\lbrace
    \begin{array}{ll}
        \varepsilon \quad&\text{if the wire is tensioned}\\ 
        x_M\varepsilon_T \quad&\text{if the wire is slacked}
    \end{array}.
    \right.
\end{equation}
Starting from a tensioned state, we initiate a slack whenever the wire begins losing tension, i.e., \(\sigma = 0\) and  \(\dot{\sigma} < 0\). When being in a slacked configuration, the wire recovers the tensioned state whenever \(\varepsilon_{eff} = \varepsilon\) and \(\dot{\varepsilon}_{eff} > 0\).

The resulting finite state machine, updated with the slack modes, is shown in Fig.~\ref{fig:hybrid2}(b). 
This hybrid automaton is composed by a set of \textit{modes} \(Q\) defined as follows
\small
\begin{equation} \label{eq:HY5}
        Q = \{Q_1, Q_2, Q_3, Q_4, Q_5\} = \{\text{AM}_0, \text{MA}_0, \text{M}_0, \text{MA}_1, \text{AM}_1\},
\end{equation}
\normalsize
and a set of \textit{edges} \(E=(Q_a,Q_b)\) representing pairs such that a transition from \(Q_a\) to \(Q_b\) is possible, i.e.,
\small
\begin{equation} \label{eq:HY6}
    \begin{aligned}
       E &= \{
        \begin{aligned}[t]
            & E_1, E_2, E_3, E_4, E_5, E_6, E_7, E_8, E_9, \\
            & E_{10}, E_{11}, E_{12}, E_{13}, E_{14}, E_{15}, E_{16}\}
        \end{aligned}\\
        &= \{ 
        \begin{aligned}[t]
            & (\text{AM}_0,\text{MA}_0), (\text{AM}_0,\text{M}_0), (\text{AM}_0,\text{AM}_1), (\text{AM}_0,\text{AM}_0), \\
            & (\text{MA}_0,\text{AM}), (\text{MA}_0,\text{MA}_1), (\text{MA}_0,\text{MA}_0), (\text{M}_0,\text{MA}_0),\\
            & (\text{M}_0,\text{MA}_1), (\text{MA}_1,\text{MA}_1),  (\text{MA}_1,\text{MA}_0), (\text{MA}_1,\text{AM}_1),\\ 
            & (\text{AM}_1,\text{AM}_1), (\text{AM}_1,\text{AM}_0), (\text{AM}_1,\text{M}_0),  (\text{AM}_1,\text{MA}_1)\}.
        \end{aligned}
    \end{aligned}
\end{equation}
\normalsize

Next, we need to define the evolution of the model states for each mode in the set \(Q\). We denote by \(\dot{\varepsilon}^{(i)}\), \(\dot{x}_M^{(i)}\), and \(\dot{T}^{(i)}\) the time derivatives of strain, phase fraction, and temperature for the generic mode \(i\). For ease of notation, we also define
\begin{align} 
    \label{eq:HY10a}
    \dot{\varepsilon}^{(i)} & \coloneqq \phi_{\varepsilon}^{(i)} = 
    \phi_{\varepsilon}^{(i)}(\varepsilon,x_M,T,v,J,T_E) \:, \\
    \label{eq:HY10b}
    \dot{x}_M^{(i)} & \coloneqq \phi_{x_M}^{(i)} = 
    \phi_{x_M}^{(i)}(\varepsilon,x_M,T,v,J,T_E) \:, \\
    \label{eq:HY10c}
    \dot{T}^{(i)} & \coloneqq \phi_{T}^{(i)} =
    \phi_{T}^{(i)}(\varepsilon,x_M,T,v,J,T_E) \:,
\end{align}
with \(i \in Q\). 
Based on \eqref{eq:MAS15}, we can compute \(\phi_{\varepsilon}^{(i)}\) and \(\phi_T^{(i)}\) for every mode without any loss of generality
\begin{equation} \label{eq:HY12}
    \phi_{\varepsilon}^{(i)} = l_0^{-1}v\:, \quad \ \forall \ i \in Q \:,
\end{equation}
\begin{equation} \label{eq:HY11a}
    \phi_T^{(i)} = \Lambda + c_V^{-1}h_{M}\phi_{x_M}^{(i)}\quad\:, \ \forall \ i \in Q \:,
\end{equation}
with
\begin{equation} \label{eq:HY11b}
    \Lambda = \left(\Omega\rho_V c_V\right)^{-1}\left[J - \lambda A_s(T-T_E)\right] \:.
\end{equation}
Instead of relying on \eqref{eq:MAS15} to also compute \(\phi_{x_M}^{(i)}\), we exploit \eqref{eq:HY2}-\eqref{eq:HY2a} to establish useful alternative relationships. When the material is transforming from A to M (\(\dot{x}_M > 0\)) and no slack occurs, we are in mode \(i = Q_1\). In here, \eqref{eq:HY2} implies
\begin{multline} \label{eq:HY14}
    \dot{\sigma}(\varepsilon,x_M) = \dot{\sigma}_A^{(n_l)}(x_M,T) \quad \Rightarrow\\
    \Rightarrow \quad 
    \pdv{\sigma(\varepsilon,x_M)}{\varepsilon}\dot{\varepsilon} + 
    \pdv{\sigma(\varepsilon,x_M)}{x_M}\dot{x}_M = \\
    = \pdv{\sigma_A^{(n_l)}(x_M,T)}{x_M}\dot{x}_M + 
    \pdv{\sigma_A^{(n_l)}(x_M,T)}{T}\dot{T} \:.
\end{multline}
By replacing \eqref{eq:HY10a}-\eqref{eq:HY11a} in \eqref{eq:HY14} and solving for \(\phi_{x_M}^{(Q_1)}\), we have
\begin{equation} \label{eq:HY15}
    \phi_{x_M}^{(Q_1)} = 
    \dfrac{\pdv{\sigma(\varepsilon,x_M)}{\varepsilon}\frac{v}{l_0} - 
    \pdv{\sigma_A^{(n_l)}(x_M,T)}{T}
    \Lambda}
    {\pdv{\sigma_A^{(n_l)}(x_M,T)}{x_M} - \pdv{\sigma(\varepsilon,x_M)}{x_M} + 
    \pdv{\sigma_A^{(n_l)}(x_M,T)}{T}\frac{h_{M}}{c_V}} \:.
\end{equation}
The partial derivatives of \(\sigma(\varepsilon,x_M)\) appearing in \eqref{eq:HY15} can be easily computed based on \eqref{eq:MAS2}. Conversely, \(\sigma_A^{(n_l)}(x_M,T)\) and \(\sigma_M^{(n_l)}(x_M,T)\) are not available analytically but rather follow from the inner loop scaling policy. Thus, one must resort to numerical methods to compute the corresponding partial derivatives, since their expression changes at each reversal point. When the SMA transforms from martensite to austenite (\(\dot{x}_M < 0\)) and no slack occurs (\(i = Q_2\)), \eqref{eq:HY2a} similarly implies
\begin{multline} \label{eq:HY14abc}
    \dot{\sigma}(\varepsilon,x_M) = \pdv{\sigma_M^{(n_l)}(x_M,T)}{x_M}\dot{x}_M + 
    \pdv{\sigma_M^{(n_l)}(x_M,T)}{T}\dot{T},
\end{multline}
and, by repeating the previous steps, we obtain
\begin{equation} \label{eq:HY15aa}
    \phi_{x_M}^{(Q_2)} = 
    \dfrac{\pdv{\sigma(\varepsilon,x_M)}{\varepsilon}\frac{v}{l_0} - 
    \pdv{\sigma_M^{(n_l)}(x_M,T)}{T}
    \Lambda}
    {\pdv{\sigma_M^{(n_l)}(x_M,T)}{x_M} - \pdv{\sigma(\varepsilon,x_M)}{x_M} + 
    \pdv{\sigma_M^{(n_l)}(x_M,T)}{T}\frac{h_{M}}{c_V}} \:.
\end{equation}
When the material is in mode \(i = Q_3\) (full martensite), \(x_M\) is constant and equal to 1, therefore  we simply have
\begin{equation} \label{eq:HY14aa}
\phi_{x_M}^{(Q_3)} = 0.
\end{equation}
During slack, identity \(\sigma = 0\) holds. This condition can be used to characterize \(\phi_{x_M}^{(i)}\) analytically for the two slack modes. For mode \(i = Q_4\), \eqref{eq:HY2} together with \(\sigma = 0\) implies
\begin{multline} \label{eq:HY16}
    \dot{\sigma}_A^{(n_l)}(x_M,T) = 0 \; \Rightarrow \\ 
    \Rightarrow \; \pdv{\sigma_A^{(n_l)}(x_M,T)}{x_M}\dot{x}_M + 
    \pdv{\sigma_A^{(n_l)}(x_M,T)}{T}\dot{T} = 0 \:,
\end{multline}
and, by replacing \(\dot{T}\) with \(\phi_T^{(i)}\) according to \eqref{eq:HY11a}, we have
\begin{equation} \label{eq:HY17}
    \phi_{x_M}^{(Q_4)} = -
    \dfrac{\pdv{\sigma_A^{(n_l)}(x_M,T)}{T}
    \Lambda}{\pdv{\sigma_A^{(n_l)}(x_M,T)}{x_M} + 
    \pdv{\sigma_A^{(n_l)}(x_M,T)}{T}\frac{h_{M}}{c_V}} \:.
\end{equation}
By using a similar reasoning for mode \(i = Q_5\), we have
\begin{equation} \label{eq:HY17a}
    \phi_{x_M}^{(Q_5)} = -
    \dfrac{\pdv{\sigma_M^{(n_l)}(x_M,T)}{T}
    \Lambda}{\pdv{\sigma_M^{(n_l)}(x_M,T)}{x_M} + 
    \pdv{\sigma_M^{(n_l)}(x_M,T)}{T}\frac{h_{M}}{c_V}} \:.
\end{equation}
As a last step, we can replace the expressions for \(\phi_{x_M}^{(i)}\) given by \eqref{eq:HY15}-\eqref{eq:HY14aa} and \eqref{eq:HY17}-\eqref{eq:HY17a} in \(\phi_T^{(i)}\), and compute in closed form the time derivative of the temperature \(\forall i \in Q\).

It is also possible to find alternative algebraic relationships to express \(x_M\) as a function of \(T\) and \(\varepsilon\). We define
\begin{equation} \label{eq:HY19}
    x_M^{(i)} \coloneqq \zeta^{(i)}_{x_M} = \zeta^{(i)}_{x_M}(\varepsilon,T) \:,
\end{equation}
representing the different phase fraction algebraic description \(\forall i \in Q\). 
By recalling \eqref{eq:HY2}-\eqref{eq:HY2a}, and since \(x_M = 1\) for \(i = Q_3\) and that \(\sigma = 0\) for \(i = Q_4\) and \(i = Q_5\), we have
\begingroup\makeatletter\def\f@size{8.5}\check@mathfonts\def\maketag@@@#1{\hbox{\m@th\normalsize\normalfont#1}}%
\begin{align}
\label{eq:HY19a}
   &\zeta^{(Q_1)}_{x_M}(\varepsilon,T) = \{x_M \in \mathbb{X}_M^{(n_l)}:\sigma_A^{(n_l)}(x_M,T) - \sigma(\varepsilon,x_M) = 0\} , \\
\label{eq:HY19b}
   &\zeta^{(Q_2)}_{x_M}(\varepsilon,T) = \{x_M \in \mathbb{X}_M^{(n_l)}:\sigma_M^{(n_l)}(x_M,T) - \sigma(\varepsilon,x_M) = 0\} , \\
\label{eq:HY19c}
   &\zeta^{(Q_3)}_{x_M} = \{1\} , \\
\label{eq:HY19d}
   &\zeta^{(Q_4)}_{x_M}(T) = \{x_M \in \mathbb{X}_M^{(n_l)}:\sigma_A^{(n_l)}(x_M,T) = 0\} , \\
\label{eq:HY19e}
   &\zeta^{(Q_5)}_{x_M}(T) = \{x_M \in \mathbb{X}_M^{(n_l)}:\sigma_M^{(n_l)}(x_M,T) = 0\} .
\end{align}\endgroup
We can use \eqref{eq:HY19}-\eqref{eq:HY19e} to eliminate any explicit dependency of \(\phi_\varepsilon^{(i)}\) and \(\phi_T^{(i)}\) on \(x_M\), thus making it no longer necessary to include the phase fraction among the continuous states of the hybrid model. As \(\sigma_A^{(n_l)}(x_M,T)\) and \(\sigma_M^{(n_l)}(x_M,T)\) are usually available in the form of look-up tables, no analytical solution is expected to be found for \(x_M\) in \eqref{eq:HY19a}-\eqref{eq:HY19e}. Therefore, for model implementation purpose, the solution of equations \eqref{eq:HY19a}-\eqref{eq:HY19e} must be addressed via numerical methods, e.g., by means of auxiliary look-up tables.
The existence of a unique solution to \eqref{eq:HY19a}-\eqref{eq:HY19e} must be ensured by proper calibration of the model parameters. As a final remark, we point out that the dependency of thermal flows $\phi_{T}^{(Q_1)}$ and $\phi_{T}^{(Q_2)}$ on the current hysteresis loop \(\sigma_A^{(n_l)}(x_M,T)\) and \(\sigma_M^{(n_l)}(x_M,T)\), in combination with the phase-dependent residual strain \eqref{eq:HY8a} occurring when the wire is in slack, represent the main mechanisms which are responsible for describing the shape memory effect.

\subsection{Hybrid Polycrystalline Model Formulation} \label{sec:PCF}
Based on the discussion in Section~\ref{sec:PC}, we can reformulate the polycrystalline SMA model \eqref{eq:MAS15} as a hybrid system \(\mathcal{H}\) in the framework of \cite{Goebel:2012} (cf. Section~\ref{sec:Hybrid}).
To this end, we consider as a continuous-time state vector
\begin{equation} \label{eq:HY3}
    x_C \coloneqq \begin{pmatrix} x_{C1}, & x_{C2} \end{pmatrix} = \begin{pmatrix} \varepsilon, & T \end{pmatrix} \in \mathbb{X}_C \:,
\end{equation}
with \(\mathbb{X}_C \coloneqq \mathbb{R}_{\geq 0} \times \mathbb{R}_{\geq 0}\), and as discrete-time state vector 
\begin{equation} \label{eq:HY4}
    x_D \coloneqq \begin{pmatrix} x_{D1}, & x_{D2}, & x_{D3} \end{pmatrix} = \begin{pmatrix} q, & s, & n_l \end{pmatrix} \in \mathbb{X}_D \:,
\end{equation}
where 
\(\mathbb{X}_D \coloneqq \{1, -1, 0\} \times \{0,1\} \times \mathbb{N}_{\geq 1}\).
Henceforth, the \(q\) values can be defined symbolically for visual simplicity as follows \(\{\text{AM}, \text{MA}, \text{M}\}\), respectively. While \(x_C\) simply accounts for the SMA strain \(\varepsilon\) and temperature \(T\), \(x_D\) holds in memory the branch typology \(q\) ($q=1$ for AM, $q=-1$ for MA, $q=0$ for M), the slack state \(s\) ($s = 0$ for no slack, $s=1$ for slack), and the current inner loop identifier \(n_l\) (according to the description provided in Section \ref{sec2}), which uniquely determine the current operative mode of the system. 
The full state vector \(x\) of the hybrid system is defined as
\begin{equation} \label{eq:HY4aaaa}
   x = \begin{pmatrix}x_C, & x_D \end{pmatrix} \in \mathbb{X}.
\end{equation}
with \(\mathbb{X} = \mathbb{X_C} \times \mathbb{X_D}\). The input vector is defined as follows
\begin{equation} \label{eq:HY7}
    u \coloneqq \begin{pmatrix} u_1, & u_2, & u_3 \end{pmatrix} = \begin{pmatrix} v, & J, & T_E \end{pmatrix} \in \mathbb{U}
\end{equation}
where $\mathbb{U}\coloneqq\mathbb{R} \times \mathbb{R}_{\geq 0} \times \mathbb{R}_{\geq 0}$.

To simplify the formalization of the four set-valued maps of the SMA hybrid model \(\mathcal{H}=(C, F, D, G)\), we define some auxiliary objects. 
These objects, denoted as $\mathcal{D}_{x_D}$, represent the jump conditions that may occur 
when the discrete state $x_D$ assumes a specific value for different operative modes $i \in Q$.
We start by defining the possible jump conditions for $q$ as
\begingroup\makeatletter\def\f@size{8.5}\check@mathfonts\def\maketag@@@#1{\hbox{\m@th\normalsize\normalfont#1}}
\begin{align} 
    \label{eq:HY31a}
    \mathcal{D}_{q1} (x,u) \coloneqq &\left\{ q = \text{AM} \colon \phi_{x_M}^{(i)} \leq 0, \: \varepsilon \leq E_M^{-1}\bar{\sigma}+\varepsilon_T \right\} , \\
    \label{eq:HY31b}
    \mathcal{D}_{q2} (x,u) \coloneqq & \left\{ q = \text{AM} \colon \varepsilon \geq E_M^{-1}\Bar{\sigma}+\varepsilon_T, \: l_0^{-1}v \geq E_M^{-1}\bar{\sigma}_S\Lambda \right\} , \\
    \label{eq:HY31c}
    \mathcal{D}_{q3} (x,u) \coloneqq & \left\{ q = \text{MA} \colon \phi_{x_M}^{(i)} \geq 0 \right\} , \\
    \label{eq:HY31d}
    \mathcal{D}_{q4} (x,u) \coloneqq & \left\{ q = \text{M} \colon \varepsilon \leq E_M^{-1}\Bar{\sigma}+\varepsilon_T, \: l_0^{-1}v \leq E_M^{-1}\bar{\sigma}_S\Lambda \right\} ,
\end{align} \endgroup
where 
\begingroup\makeatletter\def\f@size{9}\check@mathfonts\def\maketag@@@#1{\hbox{\m@th\normalsize\normalfont#1}}
\( \Bar{\sigma} \coloneqq \left.\sigma_A^{(1)}(x_M,T)\right|_{x_M = 1} \)
\endgroup
and 
\begingroup\makeatletter\def\f@size{9}\check@mathfonts\def\maketag@@@#1{\hbox{\m@th\normalsize\normalfont#1}}
\(\bar{\sigma}_S \coloneqq \left.\sigma_S(x_M)\right|_{x_M = 1}\)
\endgroup.
For instance, if the hybrid model has a discrete state $q$ = AM, the corresponding jumps are determined depending on whether the conditions given by $\mathcal{D}_{q1}$ or $\mathcal{D}_{q2}$ are met. Equivalently, the value of \(s\) produces the following slack conditions
\begingroup\makeatletter\def\f@size{8.5}\check@mathfonts\def\maketag@@@#1{\hbox{\m@th\normalsize\normalfont#1}}
\begin{align}
    \label{eq:HY32a}
    \mathcal{D}_{s1} (x,u) \coloneqq & \left\{ s = 0,\: n_l \in \mathbb{N}^e_{\geq 1} \colon \sigma_A^{(n_l)} \leq 0 , \: \phi_{\sigma_A}^{(n_l)} \leq 0 \right\} , \\
    \label{eq:HY32b}
    \mathcal{D}_{s2} (x,u) \coloneqq & \left\{ s = 0,\: n_l \in \mathbb{N}^o_{\geq 1} \colon \sigma_M^{(n_l)} \leq 0 , \: \phi_{\sigma_M}^{(n_l)} \leq 0 \right\} , \\
    \label{eq:HY32c}
    \mathcal{D}_{s3} (x,u) \coloneqq & \left\{ s = 1 \colon \varepsilon \geq \zeta_{x_M}^{(Q_4)}\varepsilon_T , \: l_0^{-1}v \geq \phi_{x_M}^{(Q_4)}\varepsilon_T \right\} , \\
    \label{eq:HY32d}
    \mathcal{D}_{s4} (x,u) \coloneqq & \left\{ s = 1 \colon \varepsilon \geq \zeta_{x_M}^{(Q_5)}\varepsilon_T , \: l_0^{-1}v \geq \phi_{x_M}^{(Q_5)}\varepsilon_T \right\} ,
\end{align} \endgroup
where \(\phi_{\sigma_A}^{(n_l)} \coloneqq \dot{\sigma}_A^{(n_l)}\) and \(\phi_{\sigma_M}^{(n_l)} \coloneqq \dot{\sigma}_M^{(n_l)}\), which can be expressed as a function of the state via \eqref{eq:HY10a}-\eqref{eq:HY10c}, \eqref{eq:HY14}, and \eqref{eq:HY14abc}. Also here, the hybrid model runs into conditions \eqref{eq:HY32c}-\eqref{eq:HY32d} or \eqref{eq:HY32a}-\eqref{eq:HY32b} depending on whether we are in slack ($s = 1$) or not ($s = 0$), respectively. Terms \(\mathbb{N}^o_{\geq k}\) and \(\mathbb{N}^e_{\geq k}\) denote the sets of odd and even integers not smaller than a given \(k\), i.e.,
\begingroup\makeatletter\def\f@size{9.2}\check@mathfonts\def\maketag@@@#1{\hbox{\m@th\normalsize\normalfont#1}}%
\begin{align} \label{eq:HY27a}
    \mathbb{N}^e_{\geq k} &\coloneqq \left\{ n\geq k \colon \exists h \in  \mathbb{N}_{\geq 1}: n = 2h - 1 \right\} , \\
    \mathbb{N}^o_{\geq k} &\coloneqq \left\{ n \geq k \colon \exists h \in  \mathbb{N}_{\geq 1}: n = 2h  \right\} .
\end{align}\endgroup
Lastly, we have jump conditions based on the range limits for $\varepsilon$ and $x_M$ in which the current branch $n_l$ is defined, given by
\begingroup\makeatletter\def\f@size{8.5}\check@mathfonts\def\maketag@@@#1{\hbox{\m@th\normalsize\normalfont#1}}%
\begin{align}
    \label{eq:HY33a}
    \mathcal{D}_{n_{l1}} (x,u) \coloneqq &
    \left\{n_l \in \mathbb{N}^e_{\geq 3} \colon \underline{x}^{(n_l)}_M \geq \zeta_{x_M}^{(i)}, \phi_{x_M}^{(i)} \leq 0 \right\} \cup \\
    & \left\{n_l \in \mathbb{N}^e_{\geq 3} \colon \zeta_{x_M}^{(i)} \geq \bar{x}^{(n_l)}_M, \phi_{x_M}^{(i)} \geq 0 \right\} \cup \\
    & \left\{n_l \in \mathbb{N}^e_{\geq 3} \colon \underline{h}^{(n_l)}_{\varepsilon_A} \geq \varepsilon_{eff}, \phi_{\varepsilon_{eff}} \leq 0 \right\} \cup \\
    & \left\{n_l \in \mathbb{N}^e_{\geq 3} \colon \varepsilon_{eff} \geq \bar{h}^{(n_l)}_{\varepsilon_A}, \phi_{\varepsilon_{eff}} \geq 0 \right\} , \\
    \label{eq:HY33b}
    \mathcal{D}_{n_{l2}} (x,u) \coloneqq & \left\{n_l \in \mathbb{N}^o_{\geq 3} \colon \underline{x}^{(n_l)}_M \geq \zeta_{x_M}^{(i)}, \phi_{x_M}^{(i)} \leq 0 \right\} \cup \\
    & \left\{n_l \in \mathbb{N}^o_{\geq 3} \colon \zeta_{x_M}^{(i)} \geq \bar{x}^{(n_l)}_M, \phi_{x_M}^{(i)} \geq 0 \right\} \cup \\
    & \left\{n_l \in \mathbb{N}^o_{\geq 3} \colon \underline{h}^{(n_l)}_{\varepsilon_M} \geq \varepsilon_{eff}, \phi_{\varepsilon_{eff}} \leq 0 \right\} \cup \\
    & \left\{n_l \in \mathbb{N}^o_{\geq 3} \colon \varepsilon_{eff} \geq \bar{h}^{(n_l)}_{\varepsilon_M}, \phi_{\varepsilon_{eff}} \geq 0 \right\} ,
\end{align}\endgroup
where 
\begingroup\makeatletter\def\f@size{9.2}\check@mathfonts\def\maketag@@@#1{\hbox{\m@th\normalsize\normalfont#1}}%
\(\underline{h}^{(n_l)}_{\varepsilon_A} = \left.h^{(n_l)}_{\varepsilon_A}\right|_{x_M = \underline{x}_M^{(n_l)}}\), \(\bar{h}^{(n_l)}_{\varepsilon_A} = \left.h^{(n_l)}_{\varepsilon_A}\right|_{x_M = \overline{x}_M^{(n_l)}}\), \(\underline{h}^{(n_l)}_{\varepsilon_M} = \left.h^{(n_l)}_{\varepsilon_M}\right|_{x_M = \underline{x}_M^{(n_l)}}\), \endgroup
and 
\begingroup\makeatletter\def\f@size{9.2}\check@mathfonts\def\maketag@@@#1{\hbox{\m@th\normalsize\normalfont#1}}%
\(\bar{h}^{(n_l)}_{\varepsilon_M} = \left.h^{(n_l)}_{\varepsilon_M}\right|_{x_M = \overline{x}_M^{(n_l)}}\), \endgroup
with
\begingroup\makeatletter\def\f@size{9.2}\check@mathfonts\def\maketag@@@#1{\hbox{\m@th\normalsize\normalfont#1}}%
\begin{align}
    h^{(n_l)}_{\varepsilon_A} = (E_M^{-1} x_M + E_A^{-1}(1 - x_M))\sigma^{(n_l)}_A(x_M,T) + \varepsilon_T x_M \:, \\
    h^{(n_l)}_{\varepsilon_M} = (E_M^{-1} x_M + E_A^{-1}(1 - x_M))\sigma^{(n_l)}_M(x_M,T) + \varepsilon_T x_M \:,
\end{align}\endgroup
while \(\varepsilon_{eff}\) is obtained by formalizing \eqref{eq:HY8a} as follows
\begingroup\makeatletter\def\f@size{9.2}\check@mathfonts\def\maketag@@@#1{\hbox{\m@th\normalsize\normalfont#1}}%
\begin{equation} \label{eq:HY29}
    \varepsilon_{eff}(x) \coloneqq \varepsilon (1 - s) + \zeta_{x_M}^{(i)}\varepsilon_T s
    \:,
\end{equation}\endgroup
and \(\phi_{\varepsilon_{eff}} = \dot{\varepsilon}_{eff}\) is given by differentiation of \eqref{eq:HY29} with \(\dot{s} = 0\)
\begingroup\makeatletter\def\f@size{9.2}\check@mathfonts\def\maketag@@@#1{\hbox{\m@th\normalsize\normalfont#1}}%
\begin{equation} \label{eq:HY34}
    \phi_{\varepsilon_{eff}} \coloneqq \dot{\varepsilon}_{eff}(x) = l^{-1}_0 v (1 - s) + \phi_{x_M}^{(i)}\varepsilon_T s
    \:.
\end{equation}\endgroup
If the state flows outside the phase fraction or strain range limits defined above, the hybrid model will jump into a new operative mode, generating the new hysteresis branch $n_l^+$ and the corresponding state variables.

According to the previous definitions, we can write the flow conditions, i.e., the ones under which the system persists in its actual condition, based on the three discrete states
\begingroup\makeatletter\def\f@size{8.5}\check@mathfonts\def\maketag@@@#1{\hbox{\m@th\normalsize\normalfont#1}}
\begin{align}
\begin{split} \label{eq:HY26a}
    \mathcal{C}_q & (x,u) \coloneqq \\
    & \left\{ q = \text{AM} \colon \phi_{x_M}^{(i)} \geq 0, \: \varepsilon \leq E_M^{-1}\Bar{\sigma} +\varepsilon_T \right\} \cup \\
    & \left\{ q = \text{MA} \colon \phi_{x_M}^{(i)} \leq 0 \right\} \cup \left\{ q = \text{M} \colon \varepsilon \geq E_M^{-1}\Bar{\sigma} +\varepsilon_T \right\},
\end{split}\\ \notag \\
\begin{split} \label{eq:HY26b}
    \mathcal{C}_s & (x,u) \coloneqq \\
    & \left\{ s = 0 , \: n_l \in \mathbb{N}^o_{\geq 1} \colon \sigma_A^{(n_l)} \geq 0 \right\} \cup \left\{ s = 1 \colon \varepsilon \leq \zeta_{x_M}^{(Q_4)}\varepsilon_T \right\} \cup \\
    & \left\{ s = 0 , \: n_l \in \mathbb{N}^e_{\geq 1} \colon \sigma_M^{(n_l)} \geq 0 \right\} \cup \left\{ s = 1 \colon \varepsilon \leq \zeta_{x_M}^{(Q_5)}\varepsilon_T \right\}.
\end{split}\\ \notag \\
\begin{split} \label{eq:HY26c}
    \mathcal{C}_{n_l} & (x,u) \coloneqq  
    \left\{ n_l < 3 \right\} \cup \\
     & \left\{ 
     n_l \in \mathbb{N}^e_{\geq 3} \colon \right.
    \begin{aligned}[t]
        \zeta_{x_M}^{(i)} \in \mathbb{X}_M^{(n_l)},  \varepsilon_{eff} \in [\underline{h}^{(n_l)}_{\varepsilon_A},\, \bar{h}^{(n_l)}_{\varepsilon_A}] \left. \right\} \cup 
    \end{aligned} \\
    & \left\{ n_l \in \mathbb{N}^o_{\geq 3} \colon \right.
    \begin{aligned}[t]
        \zeta_{x_M}^{(i)} \in \mathbb{X}_M^{(n_l)},  \varepsilon_{eff} \in [\underline{h}^{(n_l)}_{\varepsilon_M},\, \bar{h}^{(n_l)}_{\varepsilon_M}] \left. \right\} .
    \end{aligned}
\end{split}
\end{align}\endgroup
We can then define \(\mathcal{C}\) as the intersection of all conditions such that the system persists in the current operative mode
\begingroup\makeatletter\def\f@size{9}\check@mathfonts\def\maketag@@@#1{\hbox{\m@th\normalsize\normalfont#1}}
\begin{equation} \label{eq:HY30}
    \mathcal{C}(x,u) \coloneqq \mathcal{C}_q(x,u) \, \cap \, \mathcal{C}_s(x,u) \, \cap \, \mathcal{C}_{n_l}(x,u) \:, \: \: \: (x,u) \in \mathbb{X} \times \mathbb{U}.
\end{equation}\endgroup
For compactness of notation, next explicit dependency of subsets \(\mathcal{C}\) and \(\mathcal{D}\) on system states and inputs has been omitted. 

Hence, the flow set can be defined as
\begin{equation} \label{eq:HY36}
    C \coloneqq \bigcup_{m=1}^{5} C_m \:,
\end{equation}
where
\begingroup\makeatletter\def\f@size{8.7}\check@mathfonts\def\maketag@@@#1{\hbox{\m@th\normalsize\normalfont#1}}
\begin{align} \label{eq:HY37}
    C_1 &\coloneqq 
    \{(x,u) \in \mathbb{X} \times \mathbb{U} \colon q = \text{AM}, s = 0, n_{l} \in \mathbb{N}^e_{\geq 1}, x \in \mathcal{C}\}
    ,
    \\
    C_2 &\coloneqq 
    \{(x,u) \in \mathbb{X} \times \mathbb{U} \colon q = \text{MA}, s = 0, n_{l} \in \mathbb{N}^o_{\geq 1}, x \in \mathcal{C}\}
    ,
    \\
    C_3 &\coloneqq
    \{(x,u) \in \mathbb{X} \times \mathbb{U} \colon q = \text{M}, s = 0, n_{l} = 1, x \in \mathcal{C}\}
    ,
    \\
    C_4 &\coloneqq
    \{(x,u) \in \mathbb{X} \times \mathbb{U} \colon q = \text{MA}, s = 1, n_{l} \in \mathbb{N}^o_{\geq 1}, x \in \mathcal{C}\}
    ,
    \\
    C_5 &\coloneqq
    \{(x,u) \in \mathbb{X} \times \mathbb{U} \colon q = \text{AM}, s = 1, n_{l} \in \mathbb{N}^e_{\geq 1}, x \in \mathcal{C}\}
    ,
\end{align}\endgroup
and, for all \(i\in Q\), the flow map is defined as
\begin{equation}\label{eq:HY38}
    F(x, u)
    \coloneqq
    \begin{pmatrix}
        \phi_{\varepsilon}^{(i)}, & \phi_T^{(i)}, & 0, & 0, & 0 
    \end{pmatrix},\quad \forall (x, u)\in C.
\end{equation}
According to \eqref{eq:HY38}, only the strain $\varepsilon$ and temperature $T$ can undergo continuous changes during flows.
Transitions between operating modes are triggered by the following jump set
\begin{equation} \label{eq:HY39}
    D \coloneqq \bigcup_{m=1}^{16} D_m \:,
\end{equation}
where
\begingroup\makeatletter\def\f@size{8.2}\check@mathfonts\def\maketag@@@#1{\hbox{\m@th\normalsize\normalfont#1}}
\begin{align} \label{eq:HY40}
    D_1 &\coloneqq 
    \{(x,u) \in \mathbb{X} \times \mathbb{U} \colon q = \text{AM}, s = 0, n_{l} \in \mathbb{N}^e_{\geq 1}, x \in \mathcal{D}_{q1}\} ,
    \\
    D_2 &\coloneqq
    \{(x,u) \in \mathbb{X} \times \mathbb{U} \colon q = \text{AM}, s = 0, n_{l} \in \mathbb{N}^e_{\geq 1}, x \in \mathcal{D}_{q2}\} ,
    \\
    D_3 &\coloneqq
    \{(x,u) \in \mathbb{X} \times \mathbb{U} \colon q = \text{AM}, s = 0, n_{l} \in \mathbb{N}^e_{\geq 1}, x \in \mathcal{D}_{s1}\} ,
    \\
    D_4 &\coloneqq
    \{(x,u) \in \mathbb{X} \times \mathbb{U} \colon q = \text{AM}, s = 0, n_{l} \in \mathbb{N}^e_{\geq 3}, x \in \mathcal{D}_{n_{l1}}\} ,
    \\
    D_5 &\coloneqq
    \{(x,u) \in \mathbb{X} \times \mathbb{U} \colon q = \text{MA}, s = 0, n_{l} \in \mathbb{N}^o_{\geq 1}, x \in \mathcal{D}_{q3}\} ,
    \\
    D_6 &\coloneqq
    \{(x,u) \in \mathbb{X} \times \mathbb{U} \colon q = \text{MA}, s = 0, n_{l} \in \mathbb{N}^o_{\geq 1}, x \in \mathcal{D}_{s2}\} ,
    \\
    D_7 &\coloneqq 
    \{(x,u) \in \mathbb{X} \times \mathbb{U} \colon q = \text{MA}, s = 0, n_{l} \in \mathbb{N}^o_{\geq 3}, x \in \mathcal{D}_{n_{l2}}\} ,
    \\
    D_8 &\coloneqq
    \{(x,u) \in \mathbb{X} \times \mathbb{U} \colon q = \text{M}, s = 0, n_{l} = 1, x \in \mathcal{D}_{q4}\} ,
    \\
    D_9 &\coloneqq 
    \{(x,u) \in \mathbb{X} \times \mathbb{U} \colon q = \text{M}, s = 0, n_{l} = 1, x \in \mathcal{D}_{q4} \cup \mathcal{D}_{s4}\} ,
    \\
    D_{10} &\coloneqq
    \{(x,u) \in \mathbb{X} \times \mathbb{U} \colon q = \text{MA}, s = 1, n_{l} \in \mathbb{N}^o_{\geq 3}, x \in \mathcal{D}_{n_{l2}}\} ,
    \\
    D_{11} &\coloneqq
    \{(x,u) \in \mathbb{X} \times \mathbb{U} \colon q = \text{MA}, s = 1, n_{l} \in \mathbb{N}^o_{\geq 1}, x \in \mathcal{D}_{s4}\} ,
    \\
    D_{12} &\coloneqq
    \{(x,u) \in \mathbb{X} \times \mathbb{U} \colon q = \text{MA}, s = 1, n_{l} \in \mathbb{N}^o_{\geq 1}, x \in \mathcal{D}_{q3}\} ,
    \\
    D_{13} &\coloneqq
    \{(x,u) \in \mathbb{X} \times \mathbb{U} \colon q = \text{AM}, s = 1, n_{l} \in \mathbb{N}^e_{\geq 3}, x \in \mathcal{D}_{n_{l1}}\} ,
    \\
    D_{14} &\coloneqq
    \{(x,u) \in \mathbb{X} \times \mathbb{U} \colon q = \text{AM}, s = 1, n_{l} \in \mathbb{N}^e_{\geq 1}, x \in \mathcal{D}_{s3}\} ,
    \\
    D_{15} &\coloneqq 
    \{(x,u) \in \mathbb{X} \times \mathbb{U} \colon q = \text{AM}, s = 1, n_{l} \in \mathbb{N}^e_{\geq 1}, x \in \mathcal{D}_{q2}\} ,
    \\
    D_{16} &\coloneqq
    \{(x,u) \in \mathbb{X} \times \mathbb{U} \colon q = \text{AM}, s = 1, n_{l} \in \mathbb{N}^e_{\geq 1}, x \in \mathcal{D}_{q1}\} .
\end{align}\endgroup
The jump map is derived so as to enforce the transitions \(E\) previously enunciated in (\ref{eq:HY6}) and graphically shown in \figurename~\ref{fig:hybrid2}(b). In particular, we define:
\begin{equation} \label{eq:HY41}
    G(x) \coloneqq \bigcup_{m\in\{h\in E \colon x\in D_h\}} g_m(x), \quad x\in D,
\end{equation}
where:
\begingroup\makeatletter\def\f@size{8.5}\check@mathfonts\def\maketag@@@#1{\hbox{\m@th\normalsize\normalfont#1}}
\begin{align} \label{eq:HY42}
    g_1(x) &\coloneqq  
    \big(
    \begin{matrix}
        x_1, & x_2, & \text{MA}, & 0, & x_5 + 1
    \end{matrix}
    \big)&\forall x\in D_1,\\
    g_2(x) &\coloneqq 
    \big(
    \begin{matrix}
        x_1, & x_2, & \text{M}, & 0, & 1
    \end{matrix}
    \big)&\forall x\in D_2,\\
    g_3(x) &\coloneqq 
    \big(
    \begin{matrix}
        x_1, & x_2, & \text{AM}, & 1, & x_5
    \end{matrix}
    \big)&\forall x\in D_3,\\
    g_4(x) &\coloneqq 
    \big(
    \begin{matrix}
        x_1, & x_2, & \text{AM}, & 0, & x_5 - 2
    \end{matrix}
    \big)&\forall x\in D_4,\\
    g_5(x) &\coloneqq
    \big(
    \begin{matrix} \label{eq:g5}
        x_1, & x_2, & \text{AM}, & 0, & \{x_5 + 1, 1\}
    \end{matrix}
    \big)&
    \forall x\in D_5,\\
    g_6(x) &\coloneqq  
    \big(
    \begin{matrix}
        x_1, & x_2, & \text{MA}, & 1, & x_5
    \end{matrix}
    \big)
    &\forall x\in D_6,\\
    g_7(x) &\coloneqq 
    \big(
    \begin{matrix}
        x_1, & x_2, & \text{MA}, & 0, & x_5 - 2
    \end{matrix}
    \big)&\forall x\in D_7,\\
    g_8(x) &\coloneqq\big(
    \begin{matrix}
     x_1, & x_2, & \text{MA}, & 0, & 2
    \end{matrix}
    \big)&\forall x\in D_8,\\
    g_9(x) &\coloneqq  
    \big(
    \begin{matrix}
        x_1, & x_2, & \text{MA}, & 1, & 2
    \end{matrix}
    \big)&\forall x\in D_9,\\
    g_{10}(x) &\coloneqq 
    \big(
    \begin{matrix}
        x_1, & x_2, & \text{MA}, & 1, & x_5 - 2
    \end{matrix}
    \big)&\forall x\in D_{10},\\
    g_{11}(x) &\coloneqq 
    \big(
    \begin{matrix}
        x_1, & x_2, & \text{MA}, & 0, & x_5
    \end{matrix}
    \big)&\forall x\in D_{11},\\
    g_{12}(x) &\coloneqq 
    \big(
    \begin{matrix} \label{eq:g12}
        x_1, & x_2, & \text{AM}, & 1, & \{x_5 + 1, 1\}
    \end{matrix}
    \big)&\forall x\in D_{12},\\
    g_{13}(x) &\coloneqq
    \big(
    \begin{matrix}
        x_1, & x_2, & \text{AM}, & 1, & x_5 - 2
    \end{matrix}
    \big)&
    \forall x\in D_{13},\\
    g_{14}(x) &\coloneqq  
    \big(
    \begin{matrix}
        x_1, & x_2, & \text{AM}, & 0, & x_5
    \end{matrix}
    \big)
    &\forall x\in D_{14},\\
    g_{15}(x) &\coloneqq 
    \big(
    \begin{matrix}
        x_1, & x_2, & \text{M}, & 0, & 1
    \end{matrix}
    \big)&\forall x\in D_{15},\\
    g_{16}(x) &\coloneqq\big(
    \begin{matrix}
     x_1, & x_2, & \text{MA}, & 1, & x_5 + 1
    \end{matrix}
    \big)&\forall x\in D_{16}.
    \label{eq:hyb_end}
\end{align}\endgroup
so that we can designate the new operative mode in which the system will flow.
Note that the index variable \(m\) between \(D_m\) and \(g_m\) with \(m \in\{1, 2\dots,16\}\) varies following the \textit{edges} \(E\) previously defined.
In (\ref{eq:g5}) and (\ref{eq:g12}), we have a univocally determined solution that depends on \(x_M\), given by
\begin{equation}
    x_{5}^+ = \left\lbrace
    \begin{array}{ll}
        x_5 + 1 \quad& x_M > 0 \\ 
        1 \quad& x_M = 0
    \end{array}.
    \right.
\end{equation}
This allows us to determine if the model must create a new hysteresis branch (i.e., minor loop), or if it has to return to the outermost branch of the hysteresis.
Based on the defined sets and maps, the polycrystalline SMA wire can be modeled via hybrid system $\mathcal{H}=(C, F, D, G)$.

\section{Experimental validation} \label{sec4}
\noindent In this section, the SMA hybrid model is evaluated based on a large set of experimental acquisitions conducted on a quasi-plastic NiTi wire. To validate the model and demonstrate its ability in reproducing hysteresis outer/minor loops and shape memory effect, an implementation in Matlab/Simulink is carried out thanks to the \emph{Hybrid Equation (HyEQ) Toolbox} \cite{Sanfelice:2013}. 
Using the \emph{Lite HyEQ Simulator} available in the HyEQ toolbox library, it is possible to define \(\mathcal{H}=(C, F, D, G)\) functions, conditions, and hybrid states. For the numerical integration, the non-stiff Runge-Kutta solver \textit{ode45} is chosen.

\begin{figure}[!t]
\centering
\includegraphics[width=8.4cm]{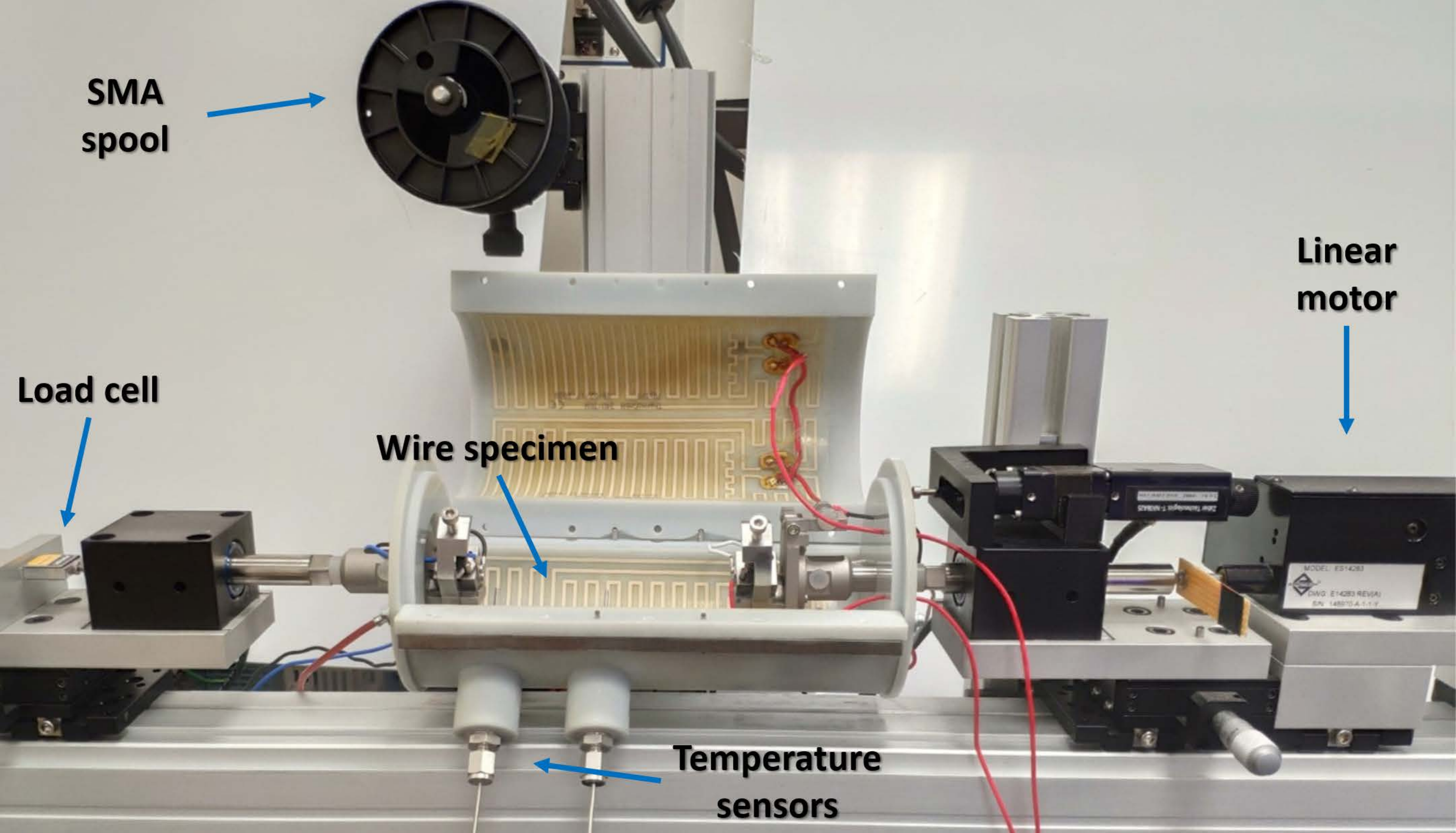}
\caption{Picture of the measurement system for the SMA wire tensile tests.}
\label{fig:results2}
\end{figure}

A total of 60 experiments are performed on a pre-trained\footnote{A thermo-mechanical training process, consisting of 100 cycles at 5 \% max strain with a power of 410 mW, is performed to stabilize the wire hysteresis.} commercial SMA wire by DYNALLOY\:, having a diameter of 75 µm and an austenitic length of 100.5 mm. The tests are conducted by varying the maximum strain, the strain rate, and the applied electrical power,  considering all the possible combinations of the following values
\begin{itemize}
    \item \textbf{Strain} 0.5 \%, 1.5 \%, 2.5 \%, 3.5 \%, 4.5 \%;
    \item \textbf{Strain rate} 0.5\(\times 10^{-3}\) s\(^{-1}\), 1\(\times 10^{-3}\) s\(^{-1}\), 5\(\times 10^{-3}\) cs\(^{-1}\);
    \item \textbf{Power} 0.5 mW, 310 mW, 360 mW, 410 mW.
\end{itemize}
Each experiment, consisting of three loading/unloading cycles, is preceded by a small (\(<\) 0.5 \%) stretching test at maximum power (500 mW). This is required to restore the material to a known initial condition, i.e., fully austenitic phase and absence of residual strain from the previous load history. 
A picture of the experimental setup is shown in \figurename~\ref{fig:results2}. The specimen under investigation is clamped at both ends and deformed constantly with triangular strain rate profiles through a linear direct drive Aerotech ANT-25LA. By means of its encoder, the linear drive acts as a displacement sensor to reconstruct the wire length, while a Futek LSB200 load cell is used to measure the SMA force. The electric power sent to the wire is regulated by a control algorithm implemented in LabVIEW. Current and voltage are measured via a NI PXI-7852R and are used to reconstruct the wire resistance. All tests are conducted in a temperature-controlled environment at 298 K. 

The parameter identification process is carried out on a restricted set of 6 experiments, corresponding to 
\begin{itemize}
    \item Strain 0.5 \%, Strain rate 0.5\(\times 10^{-3}\) s\(^{-1}\), Power 410 mW;
    \item Strain 0.5 \%, Strain rate 5\(\times 10^{-3}\) s\(^{-1}\), Power 410 mW;
    \item Strain 4.5 \%, Strain rate 0.5\(\times 10^{-3}\) s\(^{-1}\), Power 0.5 mW;
    \item Strain 4.5 \%, Strain rate 5\(\times 10^{-3}\) s\(^{-1}\), Power 0.5 mW;
    \item Strain 4.5 \%, Strain rate 0.5\(\times 10^{-3}\) s\(^{-1}\), Power 410 mW;
    \item Strain 4.5 \%, Strain rate 5\(\times 10^{-3}\) s\(^{-1}\), Power 410 mW.
\end{itemize}
The selected experiments capture the fundamental dynamics of hysteresis. All the remaining 54 experiments are used to validate the prediction of the model. Some of the model parameters are known or can be set a-priori, namely \(r_0 = 37.5 \times 10^{-6}\) m, \(l_0 = 100 \times 10^{-3}\) m, \(\rho_v = 6500\) kg/m\(^3\), \(\nu = 0.3\), and \(T_0 = 393\) K. For the polycrystalline MAS model only, we additionally set $\tau_x = 0.01$ s, $V_L = 5 \times 10^{-23}$ m$^{3}$, and $k_B = 1.38 \times 10^{-23}$ J/K. The remaining parameters are calibrated with the same procedure described in \cite{Rizzello:2019}, briefly summarized in the following. First, since it is observed that $c_V$ and $h_M$ have a negligible impact on the model output for slow experiments, all the thermo-mechanical parameters except for those two are identified based on the tests conducted with a strain rate of 0.5\(\times 10^{-3}\) s\(^{-1}\), by means of a nonlinear optimization algorithm built upon the Nelder–Mead simplex method. During this step, structural relationships are exploited to reduce the number of free parameters describing the hysteresis outer loop interpolators \eqref{eq:MAS8}-\eqref{eq:MAS10}, as in \cite{Rizzello:2019}. Subsequently, the faster experiments conducted with a strain rate of 5 \(\times 10^{-3}\) s\(^{-1}\) are used to tune $c_V$ and $h_M$. Finally, the linear dependence of the resistance on the unknown electrical parameters in \eqref{eq:MAS11}-\eqref{eq:MAS12b} is exploited to calibrate them via standard least squares. The corresponding values of the identified parameters are:
\begin{itemize}
    \item Mechanical parameters: \(E_A = 50 \times 10^{9}\) N/m\(^2\), \(E_M = 31 \times 10^{9}\) N/m\(^2\), \(\varepsilon_T = 4.07 \times 10^{-2}\);
    \item Thermal parameters: \(\lambda = 235 \) W/(K \(\times\) m\(^2\)), \(c_V = 450\) J/(K \(\times\) kg), \(h_M = 22 \times 10^{3}\) J/kg;
    \item Electrical parameters: \(\rho_{eA} = 8.11 \times 10^{-7}\) \(\Omega\) \(\times\) m, \(\rho_{eM} = 10.02 \times 10^{-7}\) \(\Omega\) \(\times\) m, \(\alpha_{eA} = 0\) 1/K, \(\alpha_{eM} = 1.4 \times 10^{-3}\) 1/K;
    \item Outer hysteresis loop parameters: \(E_{AL} = 2.397 \times 10^{8}\), \(E_{AR} = -8.765 \times 10^{8}\), \(E_{AC} = -1.560 \times 10^{9}\), \(\lambda_{AL} = 120\), \(\lambda_{AR} = 4\), \(\sigma_{AB} = 1.411 \times 10^{9}\), \(E_{ML} = 5.245 \times 10^{8}\), \(E_{MR} = -1.791 \times 10^{8}\), \(E_{MC} = -1.060 \times 10^{9}\), \(\lambda_{ML} = 9.5\), \(\lambda_{MR} = 100\), \(\sigma_{MB} = 8.263 \times 10^{8}\), \(E_{SL} = 7.709 \times 10^{6}\), \(E_{SR} = -3.904 \times 10^{5}\), \(E_{SC} = 3.997 \times 10^{5}\), \(\lambda_{SL} = 80\), \(\lambda_{SR} = 20\), \(x_{0SL} = -1.000 \times 10^{-3}\), \(x_{0SR} = 1\), \(\sigma_{SB} = 3.821 \times 10^{5}\).
\end{itemize}

\begin{figure}[!t]
\centering
\includegraphics[width=8.6cm]{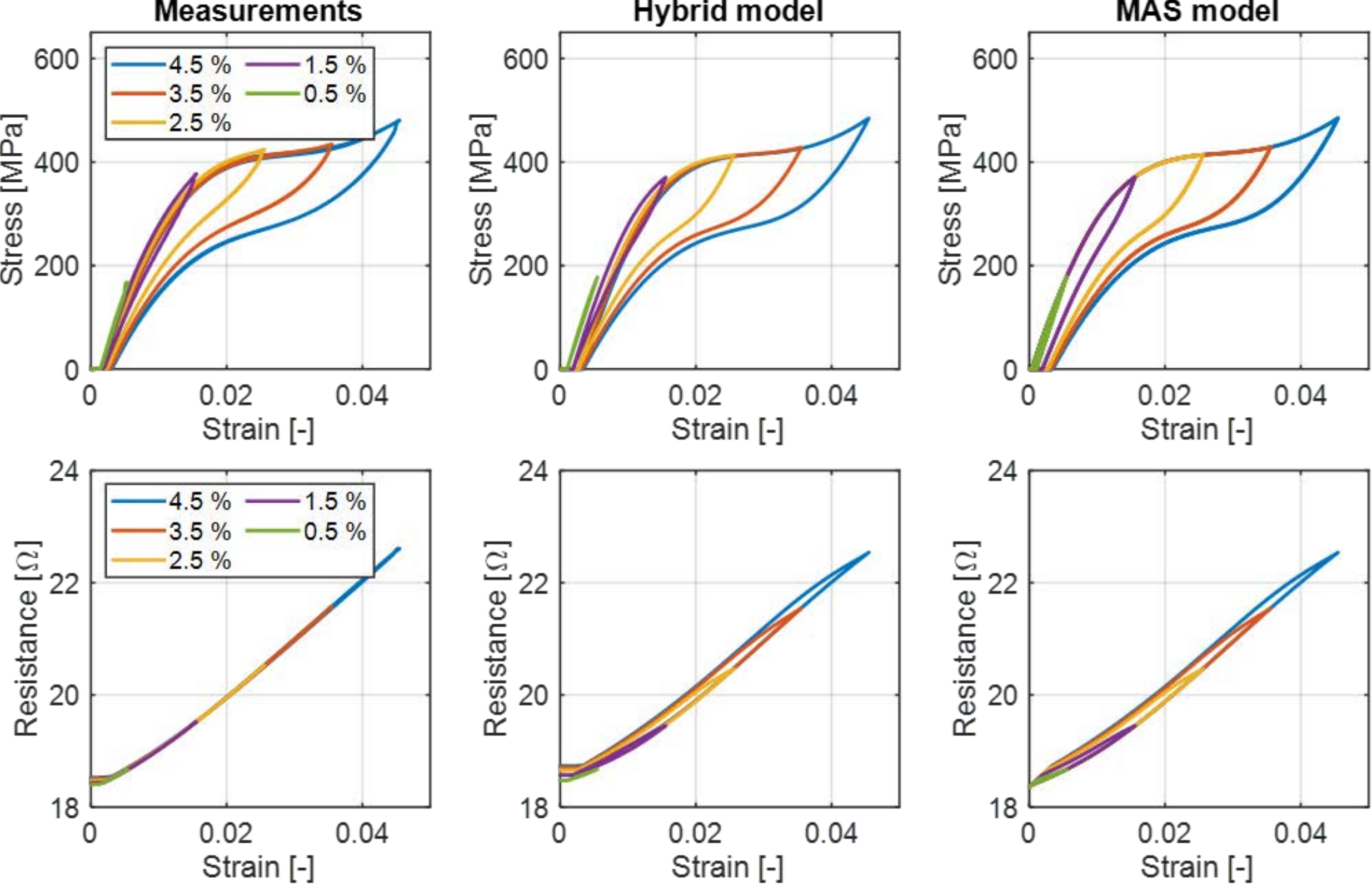}
\caption{Subset of validation experiments, obtained with a power of 360 mW and a strain rate of 0.5\(\times 10^{-3}\) s\(^{-1}\) for different strains (minor loops), is shown in the left column. Corresponding simulation results of both hybrid and MAS models are reported in the center and right columns, respectively.}
\label{fig:results3}
\end{figure}

\begin{figure*}[!t]
    \centering
    \includegraphics[width=17.9cm]{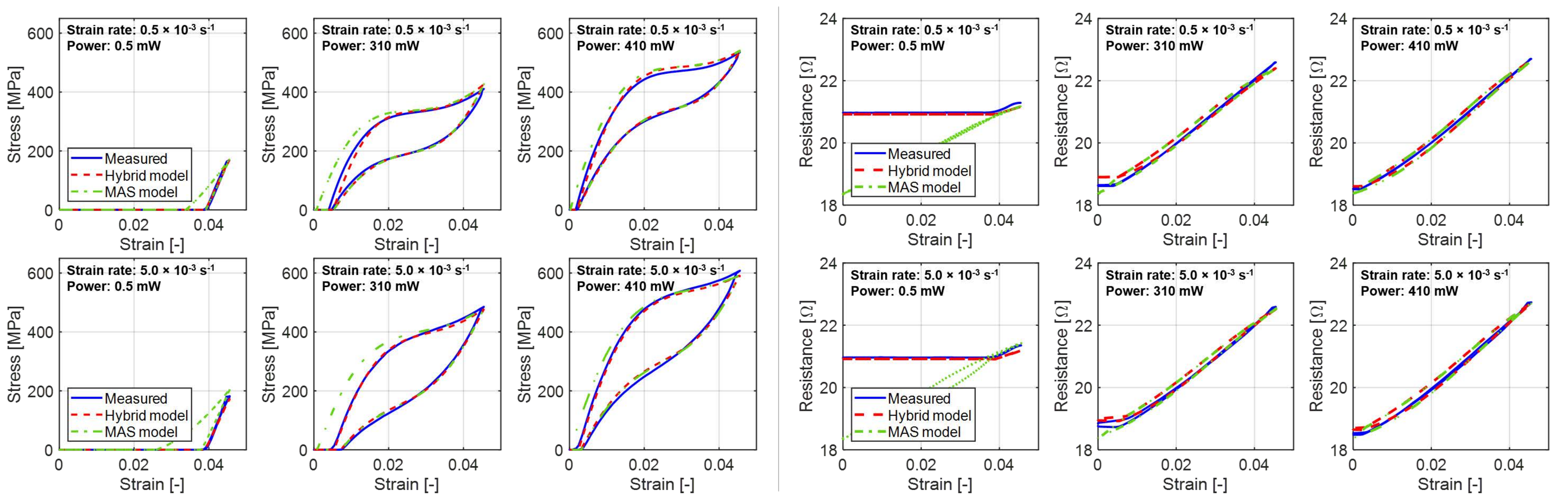}
    \caption{Subset of validation experiments with maximum strain (4.5\%). In each column, we have different powers, respectively 0.5 mW, 310 mW, and 410 mW, for stresses first and then resistances. Instead, rows differ from strain rates values, 0.5\(\times 10^{-3}\) s\(^{-1}\) and 5\(\times 10^{-3}\) s\(^{-1}\).}
    \label{fig:results4}
\end{figure*}

\begin{figure}[!t]
    \centering
    \includegraphics[width=8.5cm]{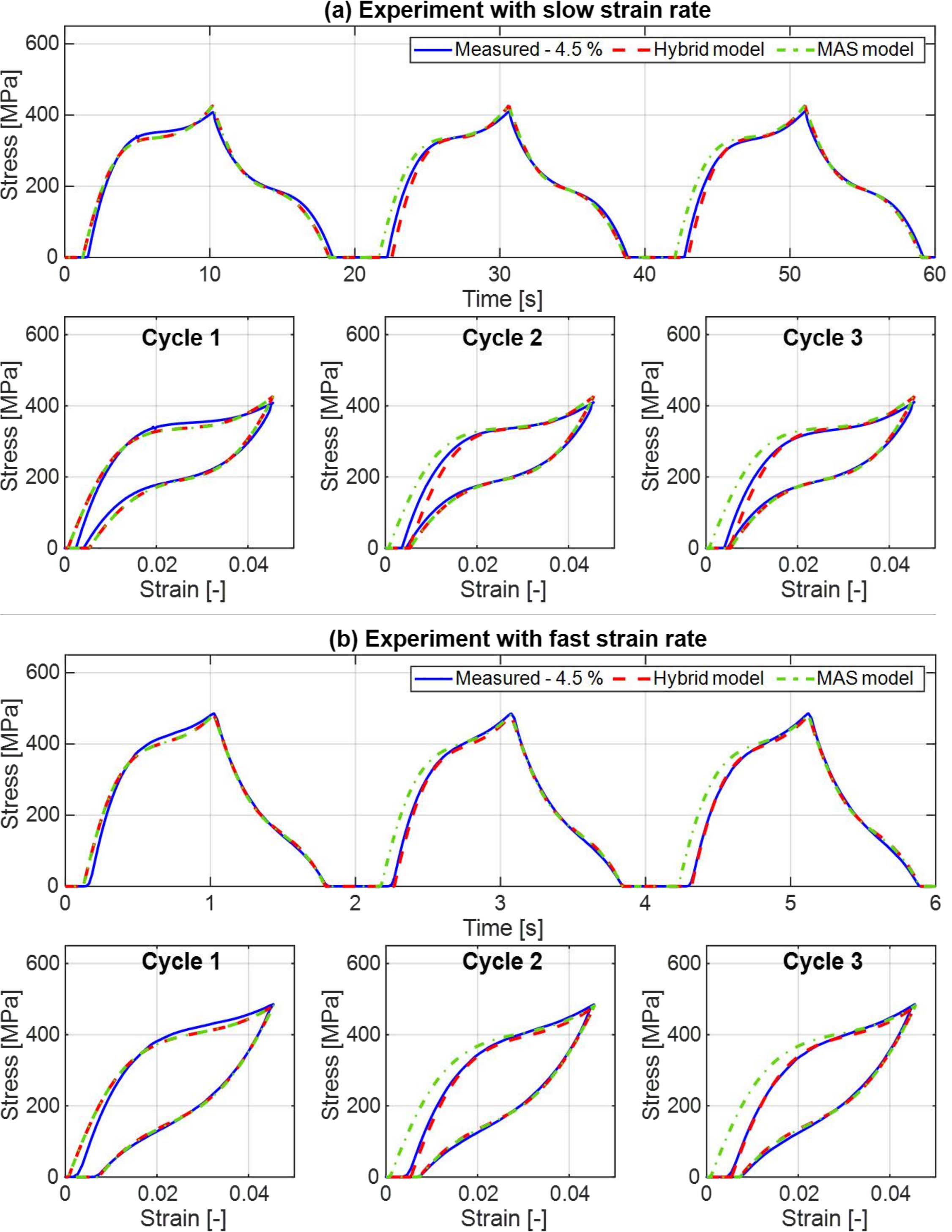}
    \caption{Comparison between the polycrystalline MAS model and the hybrid reformulation on all three cycles for two validation experiments with slow (0.5\(\times 10^{-3}\) s\(^{-1}\)) and fast (5\(\times 10^{-3}\) s\(^{-1}\)) strain rate, both corresponding to a 310 mW power.}
    \label{fig:results5}
    \vspace{-5mm}
\end{figure}

A meaningful subset of the validation results is shown in \figurename~\ref{fig:results3} and \figurename~\ref{fig:results4}, where both stress-strain and resistance-strain curves are paired together. For ease of clarity, only the final hysteresis loop is shown in those figures. From \figurename~\ref{fig:results3}, it can readily be observed that both models well reproduce all minor loops obtained experimentally for different power levels, even though calibration was performed based on minimum and maximum strain/power values only. A closer inspection of the results reveals that the hybrid model has generally higher accuracy than the polycrystalline MAS one. In \figurename~\ref{fig:results4}, it can be seen that the dependence of the hysteresis on the input power is reproduced with high fidelity for different power and strain rate values. This is especially important for SMA applications as actuators. In particular, it can be noted how the zero-stress residual strain shifts to the left for an increasing Joule heating. This phenomenon is a direct consequence of the two-way shape memory effect and is well-predicted by the proposed model. The resistance trend is overall well reproduced, as well. Note how a small hysteresis is observed in the simulated resistances in \figurename~\ref{fig:results3} and \figurename~\ref{fig:results4}, which is instead much tighter in the corresponding experimental curves. This may be due to oversimplifications in the adopted resistance model, e.g., the fact that R-phase has been neglected \cite{Wang:2021}. Also here, the hybrid model provides an overall higher accuracy than the MAS, due to the inability of the latter to predict the resistance plateaus in the low-power tests (as it lacks the slack dynamics).

The behavior of the experimental stress response during several testing cycles, starting from a zero-stress loading condition, is instead shown in \figurename~\ref{fig:results5} for a pair of selected experiments, i.e., slower (0.5\(\times 10^{-3}\) s\(^{-1}\)) and faster (5\(\times 10^{-3}\) s\(^{-1}\)) strain rate tests at 310 mW of power and 4.5 \% of maximum strain. It is clear how the residual strain causes deviations between the first and subsequent cycles. More specifically, for the slower experiment in \figurename~\ref{fig:results5}(a), the second and third loops close at a fixed residual strain which is larger than the starting value during cycle 1, as a result of the shape memory effect occurring in the material. For the faster experiment in \figurename~\ref{fig:results5}(b), instead, the wire temperature changes continuously during the slack mode, causing in turn a reduction of the residual strain between two subsequent cycles. As a result, the second and third loops no longer close at a fixed strain value. These experimental observations are automatically captured by the hybrid model, thanks to the inclusion of slack dynamics. The polycrystalline MAS model, on the other hand, incorrectly predicts that the strain returns to the starting point after each cycle, causing all loops to be identical in each experiment. To let the polycrystalline MAS model correctly describe the changing behavior of the hysteresis for different loops, a pre-processing of the input signals would be required.

A quantitative measure of the accuracy of both hybrid and polycrystalline MAS models is shown in Table~\ref{tab:results2}.
The accuracy of the models is quantified by using the same FIT index as in \cite{Rizzello:2019}, which represents a normalized root mean squared error expressed as a percentage.
Except for the smallest strain cases, in which the FIT values are small due to numerical artifacts, an accuracy higher than 90\% for the stress and 80\% for the resistance is obtained in most of the cases, thus confirming the accuracy of the hybrid model. Only in some low-power cases, the resistance fitness exhibits poor values, reported as a zero FIT in Table~\ref{tab:results2}, and resulting in an inaccurate prediction. The lower resistance accuracy is reasonably due to the presence of the small hysteresis in the simulated results, which is instead less pronounced in the corresponding experiments, see \figurename~\ref{fig:results3} and \figurename~\ref{fig:results4}. Additionally, for some experiments at low power (0.5 mW) and small strain (0.5 - 2.5 \%), no force is measured due to the wire always being in slack. Those experiments correspond to the empty cells in Table~\ref{tab:results2}. The accuracy of the polycrystalline MAS model is generally similar to the one of the hybrid model, even though its inability to predict the reversal strain during multiple cycling causes a slightly smaller accuracy overall, in agreement with what is observed in \figurename~\ref{fig:results3}-\figurename~\ref{fig:results5}.
As a further comparison, Table~\ref{tab:results2} also reports the simulation speed of both the hybrid model and the polycrystalline MAS model. For this comparison, the latter is integrated into Matlab with the stiff solver \textit{ode15s}, which is the one providing the least amount of time compared to non-stiff solvers (such as \textit{ode45}). The new hybrid model requires between 13 and 21 seconds of simulation time per test. It is observed that such simulation time mostly depends on the number of jumps since the numerical integration is fast during flows. The polycrystalline MAS model, on the other hand, exhibits simulation times similar to the hybrid one for short experiments, while this time increases by a factor of up to three for the longer experiments. This is especially visible from the last nine entries in Table~\ref{tab:results2}. The behavior can be explained considering that the polycrystalline MAS model requires stiff equation \eqref{eq:MAS13} to be integrated continuously, thus making the simulation time proportional to the experiment duration.

\begin{table}[!ht]
    \caption{Performance comparison for all the experiments.}
    \label{tab:results2}
    \centering
    \vspace{-2mm}
    \resizebox{\columnwidth}{!}{%
    \begin{tabular}{|c|c|c|c;{2pt/2pt}c|c;{2pt/2pt}c|c;{2pt/2pt}c|}
    \hline
        Max & Power & Strain & Time & Time & FIT \(\sigma\) & FIT \(\sigma\) & FIT \(R\) & FIT \(R\)  \\\relax
        Strain & Level & Rate & HYB & MAS & HYB & MAS & HYB & MAS \\\relax
        [\%] & [mW] & [\(10^{\text{-}3}\)/s] & [s] & [s] & [\%] & [\%] & [\%] & [\%] \\ \hline
        0.5 & 0.5 & 0.5 & - & - & - & - & - & - \\ \hline
        0.5 & 0.5 & 1.0 & - & - & - & - & - & - \\ \hline
        0.5 & 0.5 & 5.0 & - & - & - & - & - & - \\ \hline
        0.5 & 310 & 0.5 & 13.3 & 15.5 & 40.7 & 33.7 & 22.9 & 0.0 \\ \hline
        0.5 & 310 & 1.0 & 13.6 & 15.1 & 7.8 & 9.3 & 0.0 & 0.0 \\ \hline
        0.5 & 310 & 5.0 & 13.1 & 15.1 & 22.8 & 51.6 & 0.0 & 0.0 \\ \hline
        0.5 & 360 & 0.5 & 14.2 & 16.0 & 83.8 & 23.4 & 73.6 & 40.0 \\ \hline
        0.5 & 360 & 1.0 & 14.2 & 24.4 & 57.2 & 67.5 & 40.9 & 0.0 \\ \hline
        0.5 & 360 & 5.0 & 14.0 & 15.4 & 68.9 & 40.7 & 54.2 & 23.4 \\ \hline
        0.5 & 410 & 0.5 & 15.2 & 16.2 & 96.7 & 66.5 & 92.8 & 63.9 \\ \hline
        0.5 & 410 & 1.0 & 14.8 & 16.1 & 83.2 & 32.6 & 76.4 & 0.0 \\ \hline
        0.5 & 410 & 5.0 & 14.8 & 19.0 & 88.6 & 68.3 & 80.5 & 65.8 \\ \hline
        1.5 & 0.5 & 0.5 & - & - & - & - & - & - \\ \hline
        1.5 & 0.5 & 1.0 & - & - & - & - & - & - \\ \hline
        1.5 & 0.5 & 5.0 & - & - & - & - & - & - \\ \hline
        1.5 & 310 & 0.5 & 18.9 & 31.5 & 97.0 & 68.2 & 81.7 & 67.5 \\ \hline
        1.5 & 310 & 1.0 & 18.5 & 32.6 & 86.9 & 82.7 & 72.3 & 53.7 \\ \hline
        1.5 & 310 & 5.0 & 17.7 & 30.2 & 90.9 & 75.5 & 73.6 & 57.0 \\ \hline
        1.5 & 360 & 0.5 & 19.8 & 34.5 & 96.6 & 71.0 & 89.7 & 80.2 \\ \hline
        1.5 & 360 & 1.0 & 18.8 & 34.0 & 90.7 & 81.8 & 81.0 & 69.2 \\ \hline
        1.5 & 360 & 5.0 & 18.7 & 28.2 & 95.1 & 74.7 & 84.7 & 75.7 \\ \hline
        1.5 & 410 & 0.5 & 19.9 & 30.3 & 93.8 & 79.2 & 94.3 & 79.7 \\ \hline
        1.5 & 410 & 1.0 & 19.4 & 30.3 & 90.2 & 71.6 & 84.4 & 60.3 \\ \hline
        1.5 & 410 & 5.0 & 19.2 & 22.8 & 94.9 & 77.6 & 91.1 & 78.8 \\ \hline
        2.5 & 0.5 & 0.5 & - & - & - & - & - & - \\ \hline
        2.5 & 0.5 & 1.0 & - & - & - & - & - & - \\ \hline
        2.5 & 0.5 & 5.0 & - & - & - & - & - & - \\ \hline
        2.5 & 310 & 0.5 & 19.6 & 46.9 & 90.6 & 76.8 & 80.8 & 79.8 \\ \hline
        2.5 & 310 & 1.0 & 19.7 & 40.9 & 94.6 & 84.5 & 77.5 & 75.9 \\ \hline
        2.5 & 310 & 5.0 & 19.2 & 38.0 & 94.3 & 81.6 & 78.4 & 72.9 \\ \hline
        2.5 & 360 & 0.5 & 20.2 & 45.1 & 90.4 & 80.2 & 86.6 & 84.9 \\ \hline
        2.5 & 360 & 1.0 & 20.2 & 40.2 & 95.0 & 86.1 & 83.3 & 81.8 \\ \hline
        2.5 & 360 & 5.0 & 20.4 & 39.8 & 93.8 & 81.0 & 84.9 & 81.4 \\ \hline
        2.5 & 410 & 0.5 & 21.1 & 48.8 & 92.3 & 86.6 & 91.3 & 86.5 \\ \hline
        2.5 & 410 & 1.0 & 20.8 & 41.8 & 95.1 & 85.7 & 88.1 & 81.1 \\ \hline
        2.5 & 410 & 5.0 & 20.1 & 36.5 & 93.5 & 84.1 & 89.6 & 84.1 \\ \hline
        3.5 & 0.5 & 0.5 & 8.6 & 17.7 & 6.9 & 0.0 & 13.6 & 0.0 \\ \hline
        3.5 & 0.5 & 1.0 & 8.8 & 12.5 & 16.4 & 0.0 & 39.0 & 0.0 \\ \hline
        3.5 & 0.5 & 5.0 & 7.7 & 13.1 & 1.9 & 0.0 & 25.6 & 0.0 \\ \hline
        3.5 & 310 & 0.5 & 20.6 & 52.6 & 89.8 & 82.9 & 80.9 & 84.9 \\ \hline
        3.5 & 310 & 1.0 & 20.6 & 48.9 & 93.3 & 87.6 & 77.1 & 83.2 \\ \hline
        3.5 & 310 & 5.0 & 19.6 & 41.0 & 94.5 & 86.5 & 78.5 & 79.1 \\ \hline
        3.5 & 360 & 0.5 & 20.8 & 52.7 & 91.2 & 85.3 & 86.8 & 87.3 \\ \hline
        3.5 & 360 & 1.0 & 20.7 & 49.9 & 94.2 & 89.1 & 82.7 & 86.3 \\ \hline
        3.5 & 360 & 5.0 & 21.0 & 46.5 & 94.2 & 85.4 & 85.1 & 84.3 \\ \hline
        3.5 & 410 & 0.5 & 22.4 & 53.7 & 93.8 & 90.4 & 90.6 & 88.7 \\ \hline
        3.5 & 410 & 1.0 & 21.3 & 51.3 & 93.0 & 90.4 & 85.7 & 86.9 \\ \hline
        3.5 & 410 & 5.0 & 21.2 & 48.4 & 94.6 & 87.8 & 89.9 & 86.6 \\ \hline
        4.5 & 0.5 & 0.5 & 10.0 & 29.7 & 93.4 & 10.9 & 68.1 & 0.0 \\ \hline
        4.5 & 0.5 & 1.0 & 9.1 & 27.0 & 88.6 & 0.0 & 53.7 & 0.0 \\ \hline
        4.5 & 0.5 & 5.0 & 9.3 & 20.9 & 92.4 & 25.5 & 31.0 & 0.0 \\ \hline
        4.5 & 310 & 0.5 & 21.5 & 62.4 & 89.3 & 86.1 & 83.0 & 87.7 \\ \hline
        4.5 & 310 & 1.0 & 20.8 & 57.0 & 91.7 & 89.2 & 79.3 & 86.6 \\ \hline
        4.5 & 310 & 5.0 & 20.8 & 51.1 & 95.4 & 89.6 & 81.0 & 83.5 \\ \hline
        4.5 & 360 & 0.5 & 23.2 & 65.0 & 92.0 & 88.0 & 89.1 & 89.1 \\ \hline
        4.5 & 360 & 1.0 & 21.1 & 61.9 & 93.2 & 90.6 & 82.9 & 88.5 \\ \hline
        4.5 & 360 & 5.0 & 21.7 & 52.8 & 94.0 & 87.9 & 87.2 & 86.9 \\ \hline
        4.5 & 410 & 0.5 & 21.4 & 62.7 & 94.8 & 91.8 & 91.4 & 90.7 \\ \hline
        4.5 & 410 & 1.0 & 21.6 & 60.9 & 91.1 & 91.7 & 84.3 & 88.9 \\ \hline
        4.5 & 410 & 5.0 & 21.2 & 55.3 & 94.1 & 89.1 & 91.6 & 88.0 \\ \hline
    \end{tabular}}
    \vspace{-5mm}
\end{table}

\section{Conclusions} \label{sec5}
\noindent In this paper, a novel dynamic model is proposed to describe polycrystalline SMA wire in a lumped-parameter and control-oriented fashion for actuators. The approach relies on hybrid system theory to improve the numerical robustness and model efficiency, thus reducing simulation time without affecting physical interpretability.
Comparative studies showed high accuracy in reproducing both stress-strain and resistance-strain curves for various electro-mechanical loading histories, corresponding to different values of maximum strain, strain rate, and input power. The hybrid model well describes the complex SMA hysteresis and minor loops and also accounts for the wire behavior during slack in an automatic way.
Future studies will include model improvements, such as more accurate descriptions of the electrical resistance, or novel parameterizations of the hysteresis outer loop with the aim of reducing the number of free parameters. Furthermore, the model will be used for the simulation and dynamic optimization of SMA-driven systems, as well as for developing hybrid controllers for hysteresis compensation and self-sensing observers.
\vspace{-1mm}

\bibliographystyle{IEEEtran}
\bibliography{IEEEtran-ref}

\begin{IEEEbiography}[{\includegraphics[width=1in,height=1.25in,clip,keepaspectratio]{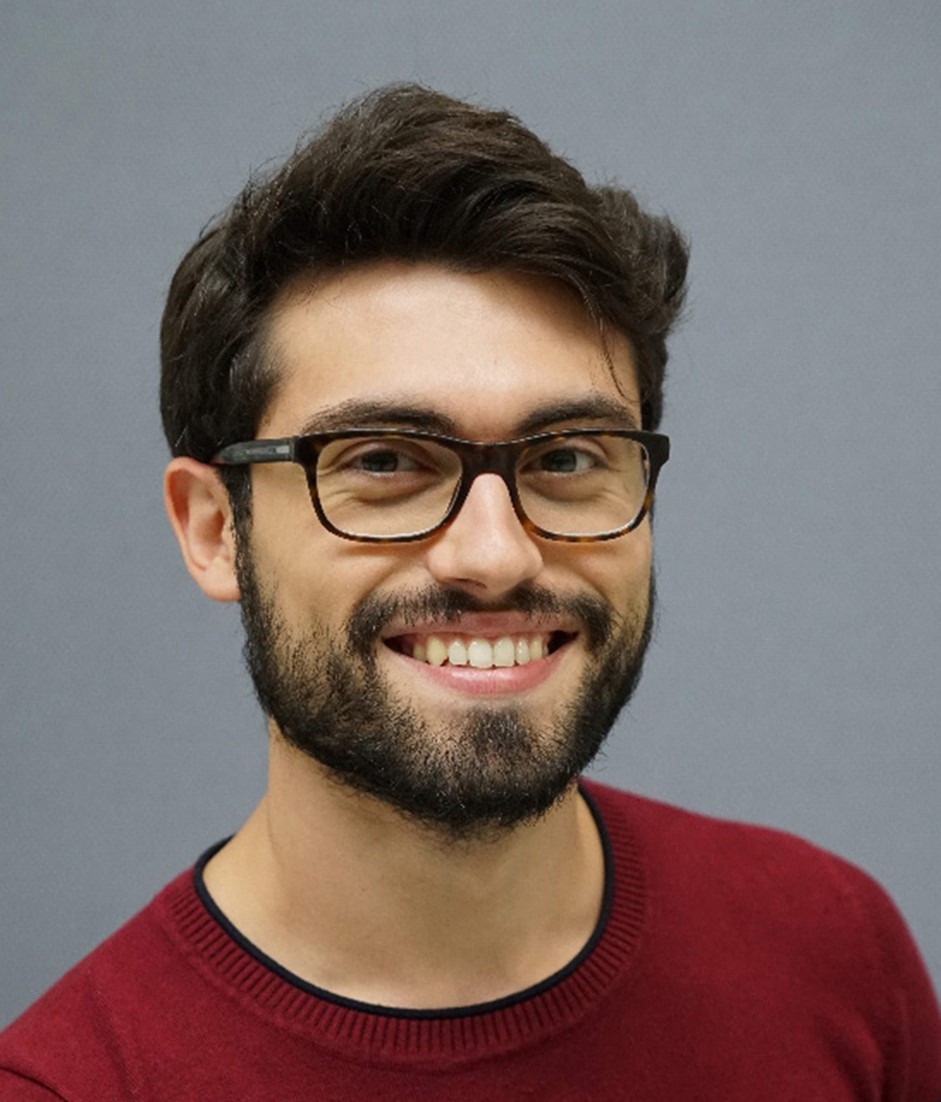}}]{Michele A. Mandolino}
was born in Gravina in Puglia, Italy, in 1994. He received B.Sc. (2015) and M.Sc. (2018) degrees in Automation and Control Engineering from the Polytechnic of Bari, Italy. He is currently a Ph.D. candidate at Saarland University under the supervision of Jun.-Prof. Dr. Gianluca Rizzello and Prof. Dr.-Ing. Stefan Seelecke. His research focuses on hybrid nonlinear modeling, control, realization, and testing of structures based on smart materials for unconventional actuators and soft robots.
\end{IEEEbiography}

\begin{IEEEbiography}[{\includegraphics[width=1in,height=1.25in,clip,keepaspectratio]{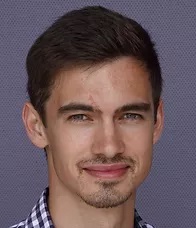}}]{Dominik Scholtes} was born 1991 in Lebach, Germany. He received is B.Eng. in Mechanical Engineering from the University of Applied Sciences in Saarbrücken in 2014 and his M.Sc. in Mechanical Engineering from Saarland University in 2017. He is currently a doctoral researcher at Saarland University under the supervision of Prof. Dr.-Ing. Stefan Seelecke. His research interests are mechanical characterization of shape memory alloy (SMA) actuator wires, joining technology of SMA micro wires as well as the development and design of actuator systems based on SMA. 
\end{IEEEbiography}

\begin{IEEEbiography}[{\includegraphics[width=1in,height=1.25in,clip,keepaspectratio]{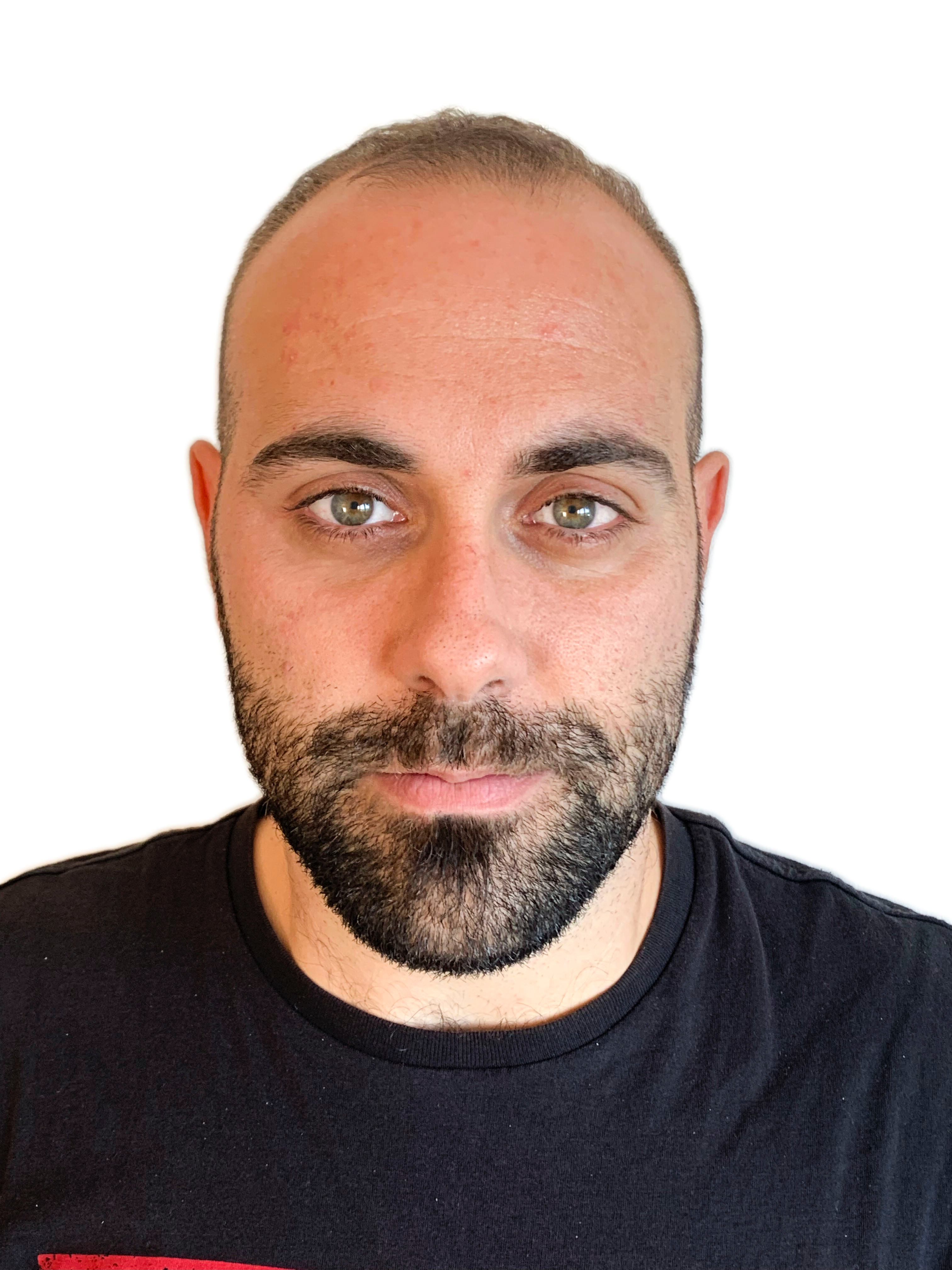}}]{Francesco Ferrante}
Francesco Ferrante (Senior Member, IEEE) received the B.Sc. (Laurea) degree in control engineering from the Sapienza – Università di Roma, Italy, in 2010, the M.Sc. (Laurea Magistrale) degree (cum laude) in control engineering from the Università degli Studi di Roma Tor Vergata, Italy, in 2012, and the Ph.D. degree in control theory from the Institut supérieur de l’aéronautique et de l’espace (SUPAERO) Toulouse, France, in 2015. From November 2015 to August 2016, he was a Post-Doctoral Fellow at the Department of Electrical and Computer Engineering, Clemson University, Clemson, SC, USA. From August 2015 to September 2016, he held a position as a Post-Doctoral Scientist at the Hybrid Systems Laboratory (HSL), University of California at Santa Cruz. From September 2017 to September 2021 he was an assistant professor at University of Grenoble Alpes, France. In September 2021 he joined the Department of Engineering of the University of Perugia, Italy where he is currently a tenure track assistant professor of systems and control. He currently serves as an Associate Editor for the IEEE Control Systems Letters, the European Journal of Control, and the IMA Journal of Mathematical Control and Information.
\end{IEEEbiography}

\begin{IEEEbiography}[{\includegraphics[width=1in,height=1.25in,clip,keepaspectratio]{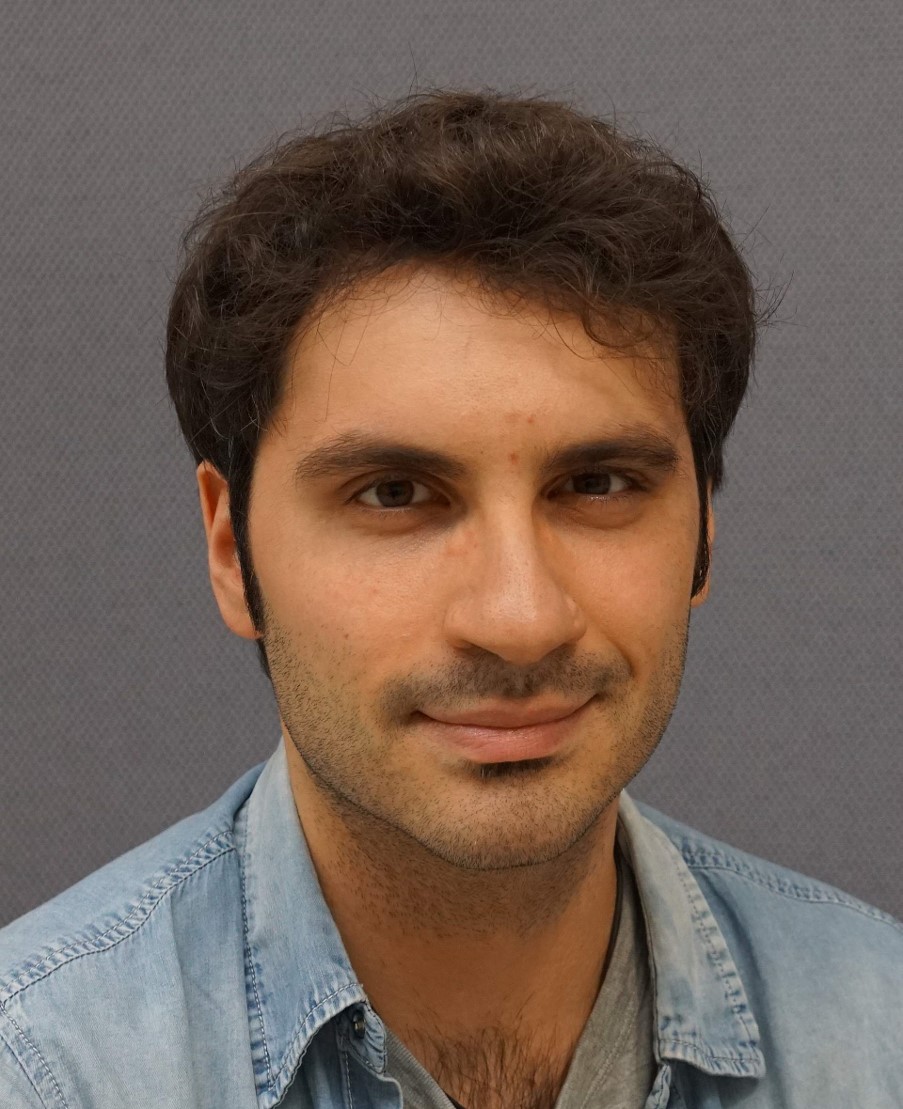}}]{Gianluca Rizzello}
(M'16) was born in Taranto, Italy, in 1987. He received the master’s (Hons.) degree in control engineering from the Polytechnic University of Bari, Bari, Italy, in 2012. He received his Ph.D. in Information and Communication Technologies from Scuola Interpolitecnica di Dottorato, a joint program between Polytechnic Universities of Torino, Bari, and Milano, Italy, in 2016. After his doctoral studies, he joined the Saarland University, Saarbrücken, Germany, first in the role of a postdoc researcher and Group Leader in Smart Material Modeling and Control (2016-2019), and subsequently as Assistant Professor in Adaptive Polymer Systems (2020 - present). His research interests involve modeling, control, and self-sensing of innovative mechatronic and robotic systems based on unconventional drive technologies, such as smart materials.

\end{IEEEbiography}

\end{document}